\documentclass[pre,twocolumn]{revtex4-1}
\usepackage{graphicx}
\usepackage{amssymb}
\usepackage{amsmath}
\begin{document}
\title{Discontinuous phase transition in chemotactic aggregation\\
with density-dependent pressure}
\author{Gyu Ho Bae}
\author{Seung Ki Baek}
\email[]{seungki@pknu.ac.kr}
\affiliation{Department of Physics, Pukyong National University, Busan 48513,
Korea}

\begin{abstract}
Many small organisms such as bacteria can attract each other by depositing chemical attractants. At the same time, they exert repulsive force on each other when crowded, which can be modeled by effective pressure as an increasing function of the organisms' density. As the chemical attraction becomes strong compared to the effective pressure, the system will undergo a phase transition from homogeneous distribution to aggregation. In this work, we describe the interplay of organisms and chemicals on a two-dimensional disk with a set of partial differential equations of the Patlak-Keller-Segel type. By analyzing its Lyapunov functional, we show that the aggregation transition occurs discontinuously, forming an aggregate near the boundary of the disk. The result can be interpreted within a thermodynamic framework by identifying the Lyapunov functional with free energy.
\end{abstract}

\maketitle

\section{Introduction}

A ubiquitous collective phenomenon in microbial and insect populations is
aggregation, which is helpful for finding food or defending themselves from
toxic
substances~\cite{shapiro1988bacteria,ma1992multicellular,shapiro1995significances,crespi2001evolution,detrain2006self,stewart2001antibiotic,secor2018entropically}.
The organisms organize their movement by depositing and sensing chemicals, which
is called chemotaxis. The canonical description of chemotactic aggregation is
provided by the Patlak-Keller-Segel (PKS)
model~\cite{patlak1953random,keller1970initiation,nanjundiah1973chemotaxis,childress1981nonlinear,stevens2000derivation}.
In two dimensions, the model predicts collapse of the whole population into a
single point when their total mass exceeds a certain threshold. Although
qualitatively correct, the prediction is physically implausible because it says
that all the mass is condensed into zero volume. To regularize this singular
behavior, a variety of mechanisms have been proposed, including
density-dependent pheromone deposition, population growth dynamics, and many
others (see Ref.~\onlinecite{hillen2009user} for a review). One of the most
convincing prescriptions is a mechanical one that incorporates the notion of
pressure~\cite{gamba2003percolation,kowalczyk2005preventing,kowalczyk2008global}.
According to this mechanism, the effective pressure between organisms increases
as they come close to each other, so aggregation ceases when their chemotactic
attraction is counterbalanced by the increased pressure. Experimentally, the
pressure in a dense insect swarm is well fitted to a polynomial function, $\Pi =
D_0 \rho + D_1 \rho^2/2! + \ldots$, where $\rho$ is the density of the organisms
and $D_i$'s are constants~\cite{sinhuber2017phase}. Also from a thermodynamic
perspective of active matter~\cite{wittkowski2014scalar,tiribocchi2015active},
such a phenomenological description of pressure holds an important place because
it can readily be defined out of equilibrium due to its mechanical
origin~\cite{takatori2014swim}, and one may integrate it over volume to derive
Helmholtz free energy by analogy to equilibrium
thermodynamics~\cite{takatori2015towards}.

In this work, we wish to take a different route toward a thermodynamic
description of chemically interacting organisms. We will consider an exact
Lyapunov functional, i.e., a nonincreasing function of
time~\cite{horstmann2001lyapunov}, for a variant of the PKS model with
density-dependent pressure~\cite{kowalczyk2005preventing,kowalczyk2008global}.
It can be identified with a free-energy functional up to an overall constant
because the long-time behavior of the system is described by the minimum of the
Lyapunov
functional~\cite{biler1998local,calvez2008parabolic,fatkullin2013study,petrosyan2014nonequilibrium,baek2017free},
just as a thermodynamic equilibrium is associated with the free-energy
minimum. We will show that this Lyapunov-functional approach gives a direct
overview of the collective dynamics, especially when the functional is expressed
in the Fourier space.

This study may also be viewed as an analysis of pattern formation in a
reaction-diffusion system, about which a vast number of references are available
(see Ref.~\onlinecite{cross1993pattern} for a review). Our model corresponds to
a multispecies reaction-diffusion
system~\cite{fanelli2013turing,madzvamuse2015cross} dealing with the
interplay of organisms and chemicals. In addition, the density-dependent
pressure $\Pi$ introduces nonlinear diffusion to this
system~\cite{gambino2014turing} as we will see below in the governing equations.
It is well known that a reaction-diffusion
system may exhibit super- or subcritical bifurcation depending on its modeling
details. One of our key research questions is whether the aggregation transition
in our model is continuous, and we will answer this question
by analyzing the Lyapunov functional.

In Sec.~\ref{sec:analysis}, we introduce our model and its Lyapunov
functional. After considering the simplest nontrivial example
to see the qualitative picture, we move on to numerical calculations
in Sec.~\ref{sec:numeric}. We discuss the numerical results as well as a
special case under radial symmetry (Sec.~\ref{sec:discussion}) and then
summarize this work in Sec.~\ref{sec:summary}.

\section{Analysis}
\label{sec:analysis}

We consider the following PKS-typed model of chemotaxis in two dimensions:
\begin{subequations}
\label{eq:kow}
\begin{align}
\frac{\partial \rho}{\partial t} &= \nabla \cdot \left(- \chi_0 \rho \nabla c +
\rho \nabla \Pi \right)
\\
\frac{\partial c}{\partial t} &= f_0 \rho + \nu_0 \nabla^2 c - g_0 c,
\label{eq:kow4}
\end{align}
\end{subequations}
where $\rho$ is the density of organisms, $c$ is the density of pheromones, and
$\Pi$ is effective pressure as a function of $\rho$.
The density dependence of pressure will be truncated at the
quadratic order, which means that
\begin{equation}
\Pi(\rho) = D_0 \rho + D_1 \rho^2/2!.
\end{equation}
As we will see below, the quadratic term introduces coupling between
modes in the simplest way.
This system contains several modeling parameters: For the organisms, the
tendency to follow a chemical gradient is given by $\chi_0$, and $D_i$'s control
how sensitively they react to a density gradient.
The second equation describes the pheromone dynamics, in which $f_0$
is the specific pheromone deposition rate, $\nu_0$ is the diffusion constant of
pheromones, and $g_0$ is the specific pheromone degradation rate.

We are dealing with a population of organisms inside a domain
$\Omega$, and the experimental timescale is assumed to be much shorter than
their typical lifetime. The total mass of the organisms, $M \equiv \int_\Omega
\rho ~dV$, must thus be conserved, where $dV$ means volume elements. This is
ensured by imposing the Neumann boundary conditions,
$\left. \nabla \rho \cdot \hat{n} \right|_{\partial\Omega} =
\left. \nabla c \cdot \hat{n} \right|_{\partial\Omega} = 0$,
where $\partial \Omega$ denotes the boundary of $\Omega$ and $\hat{n}$ is the
normal vector at an arbitrary point on $\partial \Omega$. Specifically,
we choose $\Omega$ as a two-dimensional disk of radius $l$ like a Petri dish,
whose two-dimensional volume (or area) is $V = \pi l^2$.
Note that the system of Eq.~\eqref{eq:kow} always has a
trivial homogeneous solution, $\rho = \rho_0$ and $c = (f_0 / g_0) \rho_0$,
where $\rho_0 \equiv M/V$.

Let us define $\kappa \equiv D_1/D_0$, $\lambda \equiv f_0 \chi_0 / (D_0
\nu_0)$, and $\tilde{c} \equiv (\chi_0 / D_0) c$.
In Appendix~\ref{appendix:lyapunov}, we have shown that
\begin{equation}
\begin{split}
W &\equiv \lambda \int dV \left( \kappa
\frac{\rho^3}{6} + \frac{\rho^2}{2} - \rho \tilde{c} \right)\\
&+ \int dV \left(
\frac{g_0 {\tilde{c}}^2}{2\nu_0} + \frac{|\nabla \tilde{c}|^2}{2} \right)
\end{split}
\label{eq:lyapunov}
\end{equation}
is a Lyapunov functional in the sense that $dW/dt \le 0$ (see
Ref.~\onlinecite{kowalczyk2005preventing} for the original derivation of this
functional).
To apply a variational method, it is convenient to
expand the configurations of $\rho$ and $c$ as
Fourier-Bessel series~\cite{boas2006mathematical}:
\begin{subequations}
\label{eq:expand}
\begin{align}
\rho(r,\theta,t) &= \rho_0 + \sum_{\substack{p=0\\m=1}}^{\infty} Q_{pm}
J_p(j'_{pm}r/l) \cos [p(\theta-\eta_{pm})] \label{eq:rho_expand}\\
c(r,\theta,t) &= c_0 + \sum_{\substack{p=0\\m=1}}^\infty R_{pm}
J_p(j'_{pm}r/l) \cos [p(\theta-\phi_{pm})] \label{eq:c_expand},
\end{align}
\end{subequations}
where $J_p$ is the Bessel function of the first kind and
we have defined $j'_{pm}$ as the $m$th zero
of $\frac{d}{dx} J_p(x) = 0$ to satisfy the boundary conditions.
For example, $j'_{11} \approx 1.841$, $j'_{21} \approx 3.054$ and $j'_{01}
\approx 3.832$ are the smallest ones among all the possible
cases of $p$ and $m$.
Let us denote the set of time-dependent coefficients and phases as
$\mathcal{S} \equiv \{ c_0, Q_{pm}, R_{pm}, \eta_{pm}, \phi_{pm} | p=0,\ldots,
\infty, m=1,\ldots, \infty \}$.
Its elements change over time with decreasing $W$, and trajectory must be
confined in a physical region in which both $\rho$ and $c$ are non-negative.
We wish to find $\mathcal{S}$ that minimizes Eq.~\eqref{eq:lyapunov} inside the
physical region, expecting that it will describe stationary behavior in the long
time. The result will be compared with direction integration of
Eq.~\eqref{eq:kow}.

If $\kappa=0$, then pressure is linearly proportional to $\rho$, and we can
calculate Eq.~\eqref{eq:lyapunov} in a closed form as follows (see
Appendix~\ref{sec:series}):
\begin{eqnarray}
\frac{W_\text{st}^{(\kappa=0)}}{V} &=& \frac{\lambda}{4}
\sum_{p=0}^\infty
\sum_{m=1}^\infty
(1+\delta_{p0}) J_p^2(j'_{pm}) \left( 1-\frac{p^2}{j'^2_{pm}} \right)
z_{pm}\nonumber\\
&&\times
\left( z_{pm} - \frac{\chi_0}{D_0} \right) R_{pm}^2
+ \frac{W_\text{hom}}{V},
\label{eq:quad}
\end{eqnarray}
where the subscript ``st'' on the left-hand side means stationarity, i.e.,
$\partial \rho/ \partial t = \partial c/\partial t=0$, and $W_\text{hom}$ is
contribution from the homogeneous solution.
We have also introduced in Eq.~\eqref{eq:quad} a composite parameter,
\begin{equation}
z_{pm} \equiv \frac{\nu_0 j'^2_{pm}}{f_0 l^2} + \frac{g_0}{f_0}.
\label{eq:zpm}
\end{equation}
Equation~\eqref{eq:zpm} contains $f_0 l^2 / \nu_0$, implying
competition between the timescales of pheromone deposition and
diffusion, and the second term is a ratio between the specific degradation and
deposition rates of pheromones.
Two points are clear from Eq.~\eqref{eq:quad}:
First, the Lyapunov functional for $\kappa=0$
is exactly decomposed into contributions of
separate modes. Second, each contribution is a quadratic function of
$R_{pm}$'s: If $z_{pm} < \chi_0/D_0$, then the corresponding coefficient
$R_{pm}$ can grow with decreasing the Lyapunov functional.
When the chemotactic coupling strength gradually increases, therefore,
the first excited mode is the one that has the lowest $j'_{pm}$, which is
$j'_{11} \approx 1.841$. The relaxation timescale diverges at
this point because the landscape of Eq.~\eqref{eq:quad} is flat along the
$R_{11}$ axis. When $z_{11} < \chi_0 / D_0$,
the excitation of $R_{11}$ is bounded only by
the condition that both $\rho$ and $c$ must be non-negative. Thus, if we define
an order parameter of aggregation as
\begin{equation}
A \equiv \max_{\vec{r}} \rho(\vec{r}) - \min_{\vec{r}} \rho(\vec{r}),
\label{eq:order}
\end{equation}
then it shows a jump at the threshold. However, differently from a usual
discontinuous transition, no hysteresis exists in this picture.

\begin{figure}
\includegraphics[width=0.45\columnwidth]{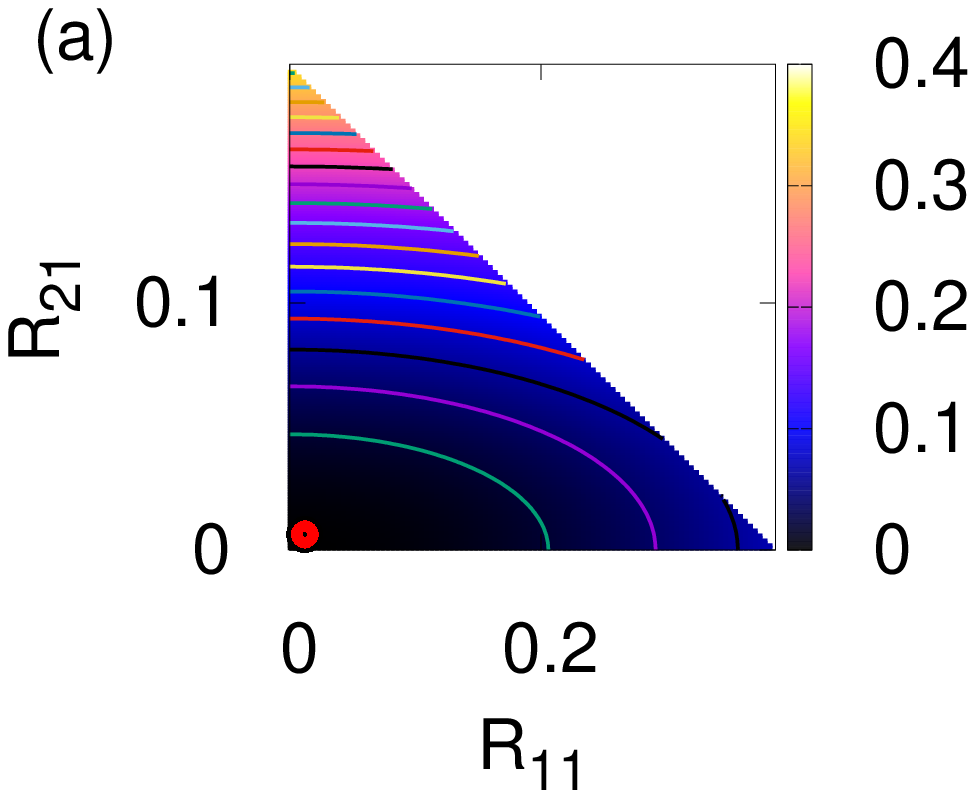}
\includegraphics[width=0.45\columnwidth]{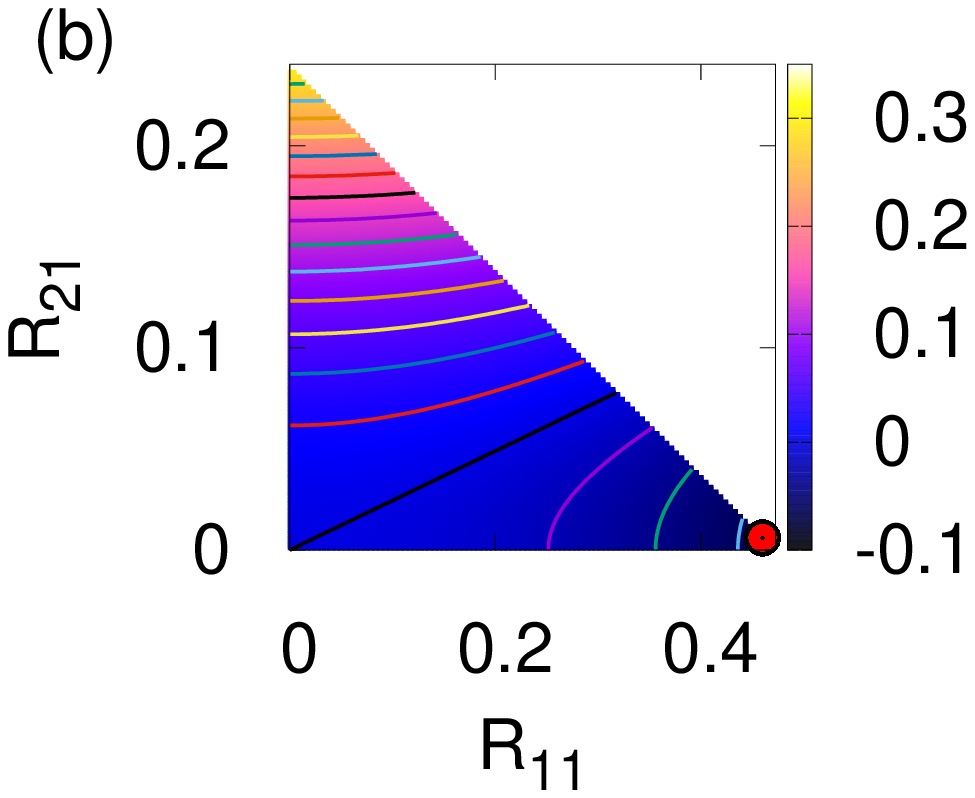}
\includegraphics[width=0.45\columnwidth]{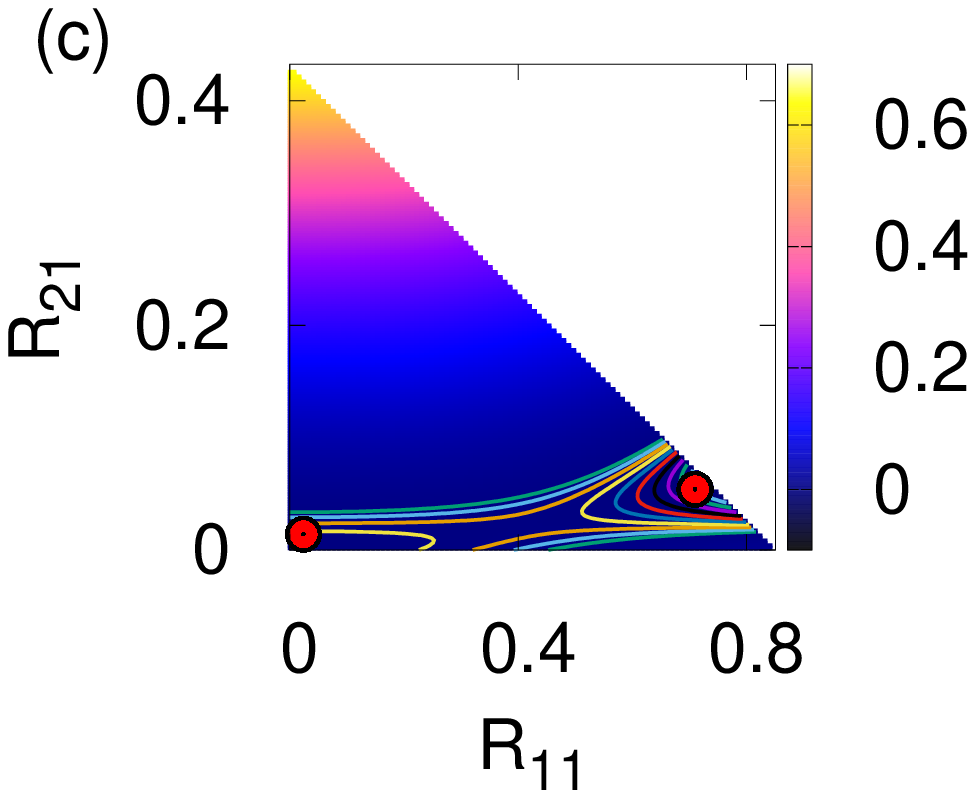}
\includegraphics[width=0.45\columnwidth]{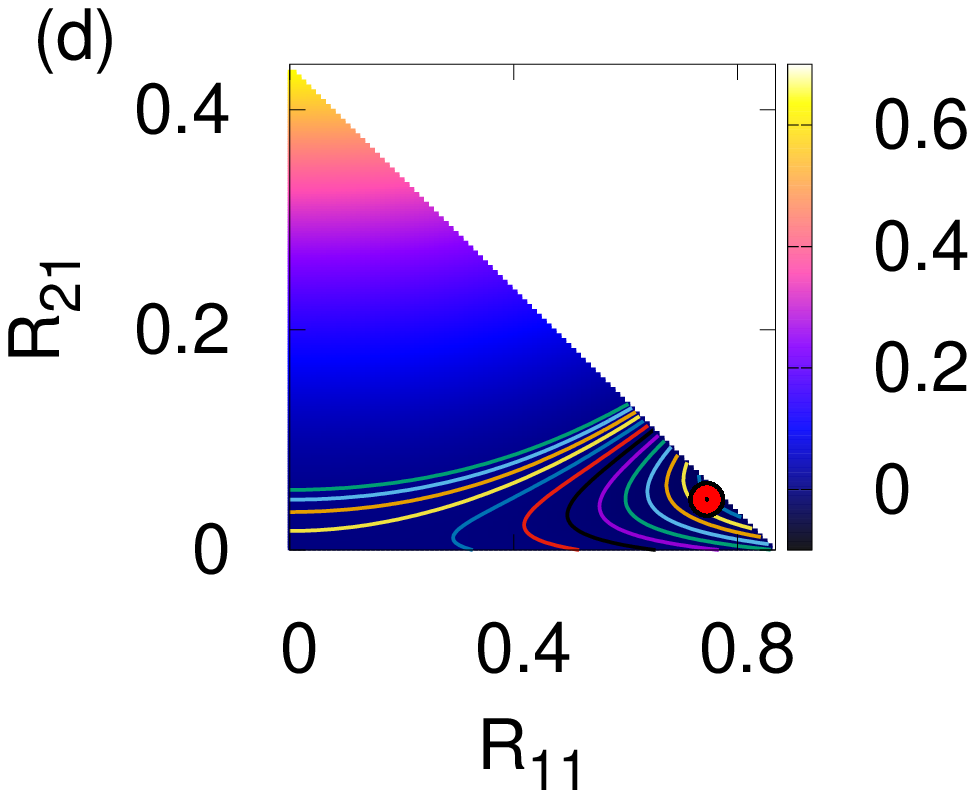}
\caption{Schematic representation of $W_\text{st}$ under the two-mode
approximation [Eq.~\eqref{eq:double}] to illustrate how aggregation transition
occurs as the strength of chemotaxis grows. The color code means the value of
Eq.~\eqref{eq:double} in arbitrary units.
The red dots denote local minima,
and we are plotting the region in which both $\rho$ and $c$ are non-negative.
(a) If $\kappa=0$, then the origin is the only minimum below the transition
point, and (b) when the system undergoes the transition,
$R_{11}$ jumps to the maximal possible value while $R_{21}$ remains zero.
(c) If $\kappa>0$, then the coupling term forms another local
minimum in addition to the origin,
before the homogeneous solution loses stability.
(d) When the aggregation transition has occurred,
$R_{21}$ may also take a nonzero value due to the coupling.
}
\label{fig:xy}
\end{figure}

Now, we turn on $\kappa$ in Eq.~\eqref{eq:lyapunov} to introduce
coupling between modes. By way of illustration,
let us take the two lowest modes and calculate the Lyapunov functional for
$\kappa > 0$:
\begin{eqnarray}
\frac{W_\text{st}}{V}
&\approx& \frac{\lambda}{4} J_1^2(j'_{11})
\left(1-\frac{1}{j'^2_{11}}\right) z_{11} \left[ (1+\kappa\rho_0)
z_{11} - \frac{\chi_0}{D_0} \right] R_{11}^2 \nonumber\\
&+&\frac{\lambda}{4} J_2^2(j'_{21})
\left(1-\frac{4}{j'^2_{21}}\right) z_{21} \left[ (1+\kappa\rho_0)
z_{21} - \frac{\chi_0}{D_0} \right] R_{21}^2 \nonumber\\
&+& \kappa\frac{\lambda}{4} I_{112} \cos2(\eta_{11}-\eta_{21}) z_{11}^2
z_{21} R_{11}^2 R_{21}
+ \frac{W_\text{hom}}{V},
\label{eq:double}
\end{eqnarray}
where $I_{112} \equiv \int_0^1 x J_1^2(j'_{11}x) J_2(j'_{21}x) dx \approx
0.0474258$.
Equation~\eqref{eq:double} contains a combination of parameters,
$(1+\kappa \rho_0) z_{11} - \chi_0 / D_0$, which crucially determines
the landscape of $W_\text{st}$. Note that it is negatively proportional
to $\lambda \equiv f_0 \chi_0 / (D_0 \nu_0)$, a ratio between chemotactic
coupling strength ($f_0 \chi_0$) and diffusivity ($D_0 \nu_0$).
The coefficient of $R_{11}^2$ vanishes when this combination equals zero, or
equivalently, when
\begin{equation}
z_{11} = \frac{\chi_0 / D_0}{1+ \kappa \rho_0}.
\label{eq:crit}
\end{equation}
This coincides with a prediction of linear stability analysis
(Appendix~\ref{appendix:linear}).
To see an effect of the coupling term $\propto \kappa$, let us assume
that $R_{11} > 0$. As we have seen above,
if $\kappa$ was zero, then $R_{21}$ could not be excited because
it would only increase $W_\text{st}$ due to
the second term on the right-hand side of Eq.~\eqref{eq:double}.
However, when $\kappa\neq 0$,
the coupling effect may compete with the second term: If the coupling
decreases $W_\text{st}$ enough to compensate for the second term, then
$R_{21}$ will also be excited. As we include more and more modes in
Eq.~\eqref{eq:double}, we will see a cascade of excitation among them,
which is essentially switched on by $R_{11}$.

The question is then whether such coupling can render the transition continuous.
The answer is negative, at least within this two-mode approximation.
Let us visualize the landscape of Eq.~\eqref{eq:double}.
We may regard $R_{pm}$'s as positive if $p>0$, and
the phase factor $\eta_{11}$ can be set to zero without loss of generality.
We may also choose $\eta_{21} = \pi/2$ to make the coupling term reduce
$W_\text{st}$ to the fullest extent. Then, $W_\text{st}$ becomes a function of
$R_{11}$ and $R_{21}$, and its landscape can readily be depicted as in
Fig.~\ref{fig:xy}. Suppose $\kappa=0$ as an example:
$R_{11}$ will exhibit an abrupt
jump at the aggregation transition as the strength of chemotaxis grows, while
$R_{21}$ remains zero [Fig.~\ref{fig:xy}(a) and (b)]. Unlike a usual
discontinuous transition, however, the minimum of $W^{(\kappa=0)}_\text{st}$ is
always unique as pointed out below Eq.~\eqref{eq:order}.
When $\kappa \neq 0$, on the other hand, multiple local minima may coexist
in the vicinity of the aggregation transition [Fig.~\ref{fig:xy}(c)], although
the aggregate will contain more than one wavenumbers as we see $R_{21} > 0$ in
Fig.~\ref{fig:xy}(d). Such bistability is a characteristic feature of a
discontinuous transition.

\section{Numerical calculation}
\label{sec:numeric}

\subsection{Gradient descent}
\begin{figure}
\includegraphics[width=0.49\columnwidth]{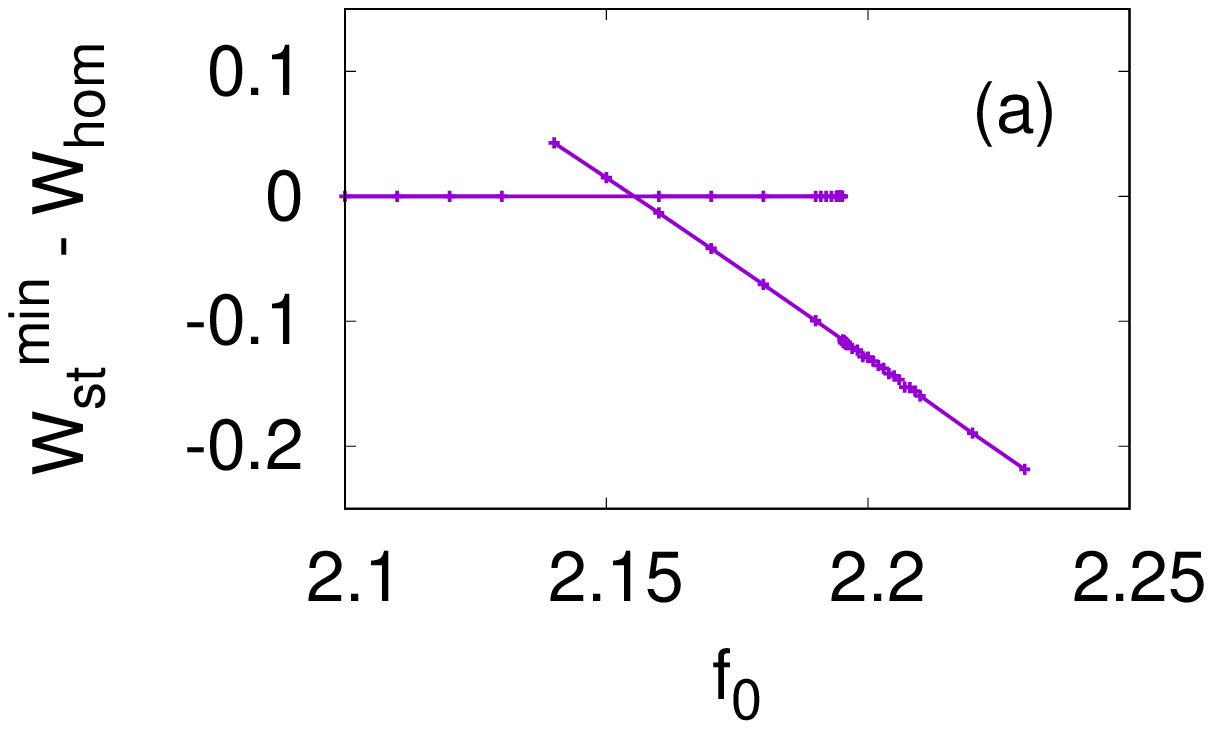}
\includegraphics[width=0.49\columnwidth]{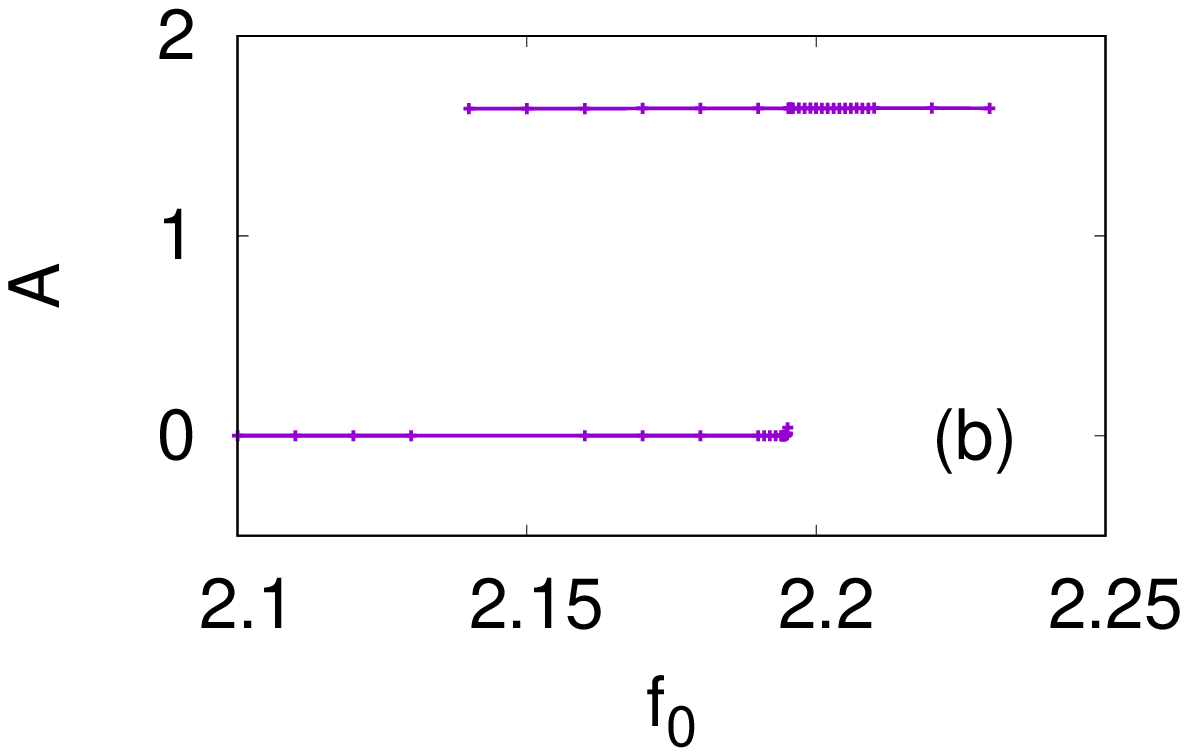}
\includegraphics[width=0.49\columnwidth]{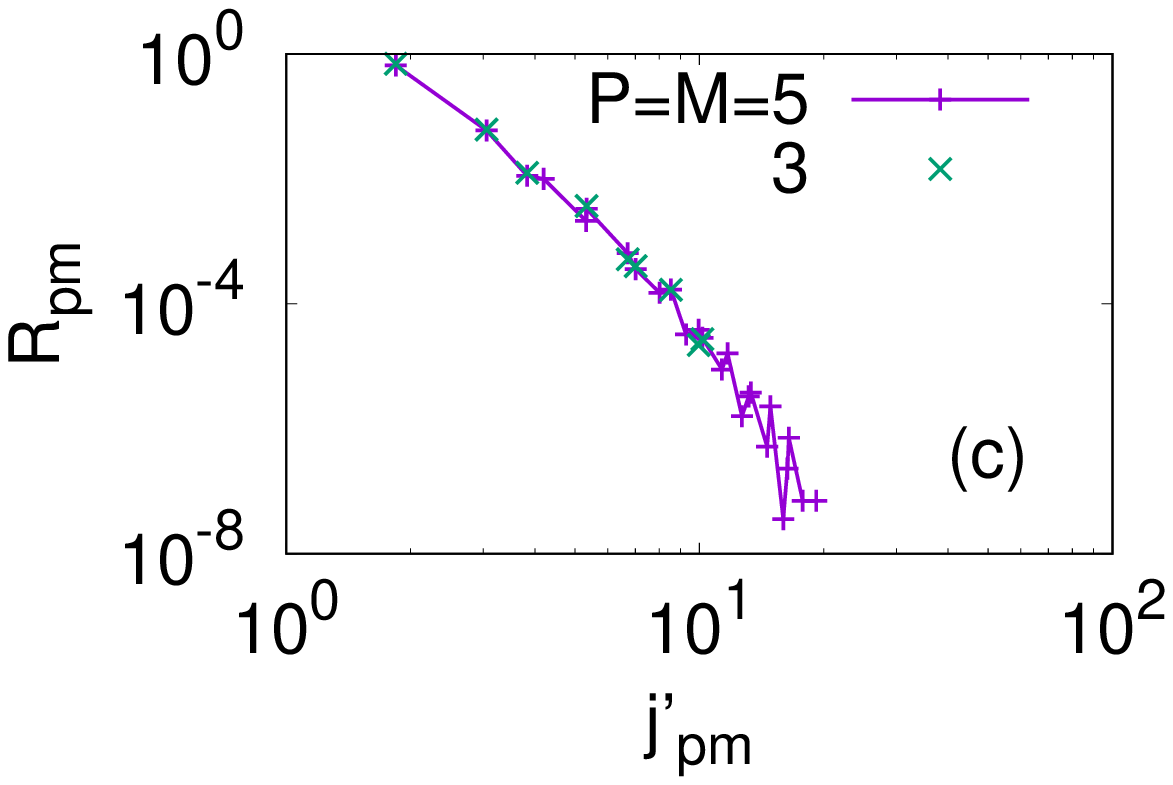}
\includegraphics[width=0.49\columnwidth]{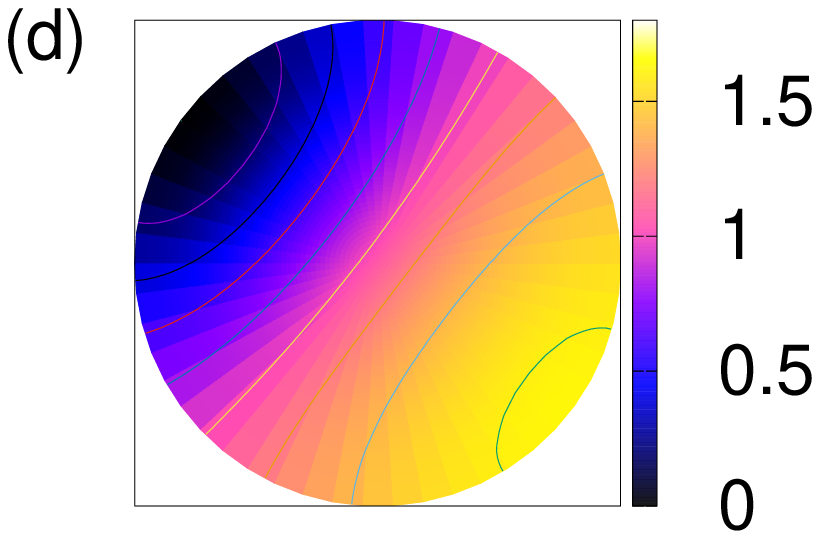}
\caption{(a) Aggregational part of the Lyapunov functional, i.e., $W^\text{min}
_\text{st} - W_\text{hom}$, and (b) the order parameter [Eq.~\eqref{eq:order}]
as functions of $f_0$. As indicated by the superscript ``min'',
these are obtained
by numerical minimization of the approximate Lyapunov functional with $P=M=3$.
(c) Coefficients of the modes when aggregation has occurred, and (d) the
corresponding density profile $\rho(r,\theta)$.
We have set $\chi_0=4$ and $\rho_0 = l = D_0 = \nu_0 = g_0 = \kappa = 1$ in this
calculation.
}
\label{fig:grad}
\end{figure}

To see how the above two-mode picture generalizes to a multi-mode case,
we perform numerical calculation. The result
confirms that the two-mode picture has already captured the
essential physics of aggregation transition:
Let us begin by approximating $W_\text{st}$ by including modes with
$p<P$ and $m \le
M$ in Eq.~\eqref{eq:expand}. The approximate
functional is then minimized by the gradient descent method with momentum,
\begin{equation}
\Delta \vec{w}' = - \gamma \nabla W_\text{st} + \alpha \Delta \vec{w},
\end{equation}
where $\vec{w}$ denotes a $(2PM)$-dimensional vector of $E_{pm} \equiv R_{pm}
\cos p\phi_{pm}$ and $F_{pm} \equiv R_{pm} \sin p\phi_{pm}$, and the prime on
the left-hand side means an updated quantity.
We have chosen the learning rate $\gamma = 10^{-5}$ and the momentum rate
$\alpha = 0.9$.
Note that $\rho$ and $c$ may become negative in
minimizing $W_\text{st}$, whereas the non-negativity is guaranteed by
the original dynamics,
Eq.~\eqref{eq:kow}~(see
Ref.~\onlinecite{kowalczyk2005preventing} for a proof). An update of $\vec{w}$
must thus be permitted only when it keeps $\rho$ and $c$ non-negative.
Specifically, we have checked the effect of updating $\vec{w}$ componentwise:
A component will preserve its old value if its update violates the
non-negativity condition.
If no components of $\vec{w}$ can be updated under this
condition, the calculation is terminated. We have also repeated this
calculation with a smaller value of $\gamma$ and found consistent results.
Following Ref.~\onlinecite{baek2017free}, we set $\chi_0=4$ and choose
other parameters, $\rho_0$, $l$, $D_0$, $\nu_0$, $g_0$, and $\kappa$, as unity.
Our control parameter is the specific pheromone deposition rate $f_0$.
According to Eq.~\eqref{eq:crit},
the homogeneous solution loses stability as $f_0$ exceeds
$f_0^\text{unstable} \approx 2.195$.
Although our calculation has used simple parameter values to focus on the
physical mechanism, interested readers may find some of their experimental
estimates for {\it Escherichia coli} in
Refs.~\onlinecite{ford1991measurement,saragosti2011directional}.

The minimization results clearly show that the system undergoes
a discontinuous transition [Figs.~\ref{fig:grad}(a) and \ref{fig:grad}(b)].
The two branches are obtained by starting the gradient descent algorithm from
different initial conditions. The crossing point of the branches
in Fig.~\ref{fig:grad}(a) indicates where the transition occurs. Once
$R_{11}$ becomes nonzero, the other modes also get excited
[Fig.~\ref{fig:grad}(c)]. However, the amplitudes are narrowly distributed, so
the aggregate has compact geometry [Fig.~\ref{fig:grad}(d)].

\subsection{Spectral method}

\begin{figure}
\includegraphics[width=0.49\columnwidth]{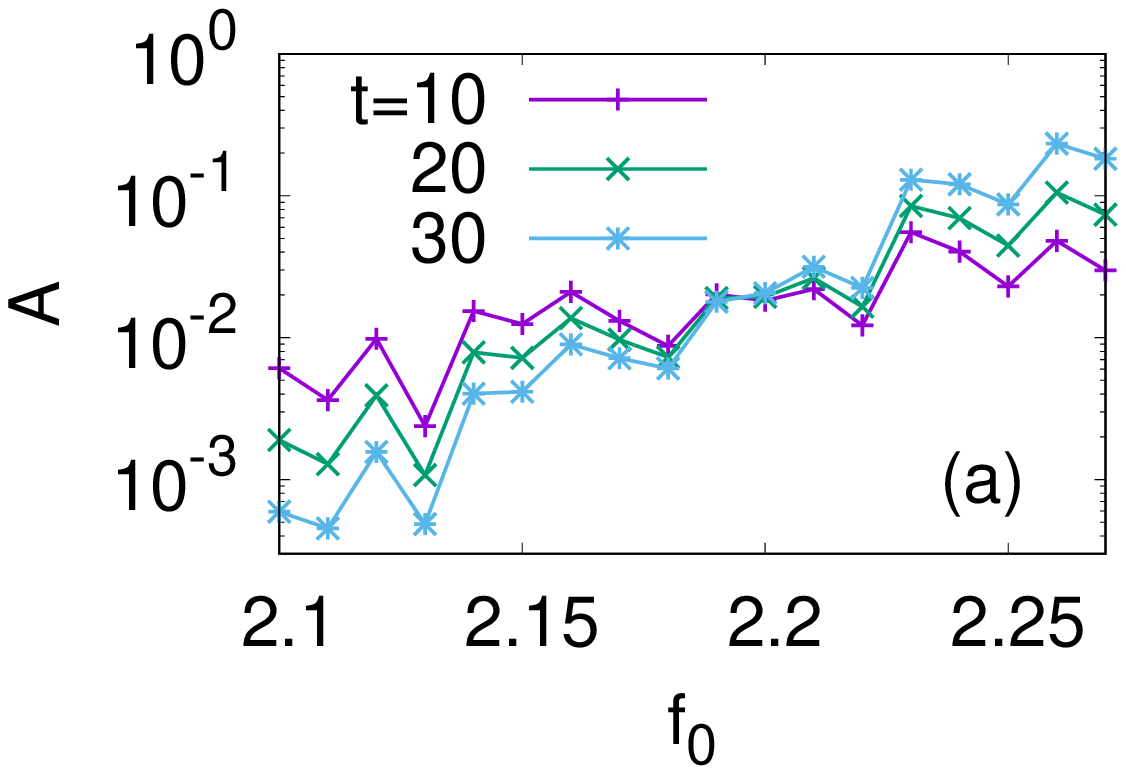}
\includegraphics[width=0.49\columnwidth]{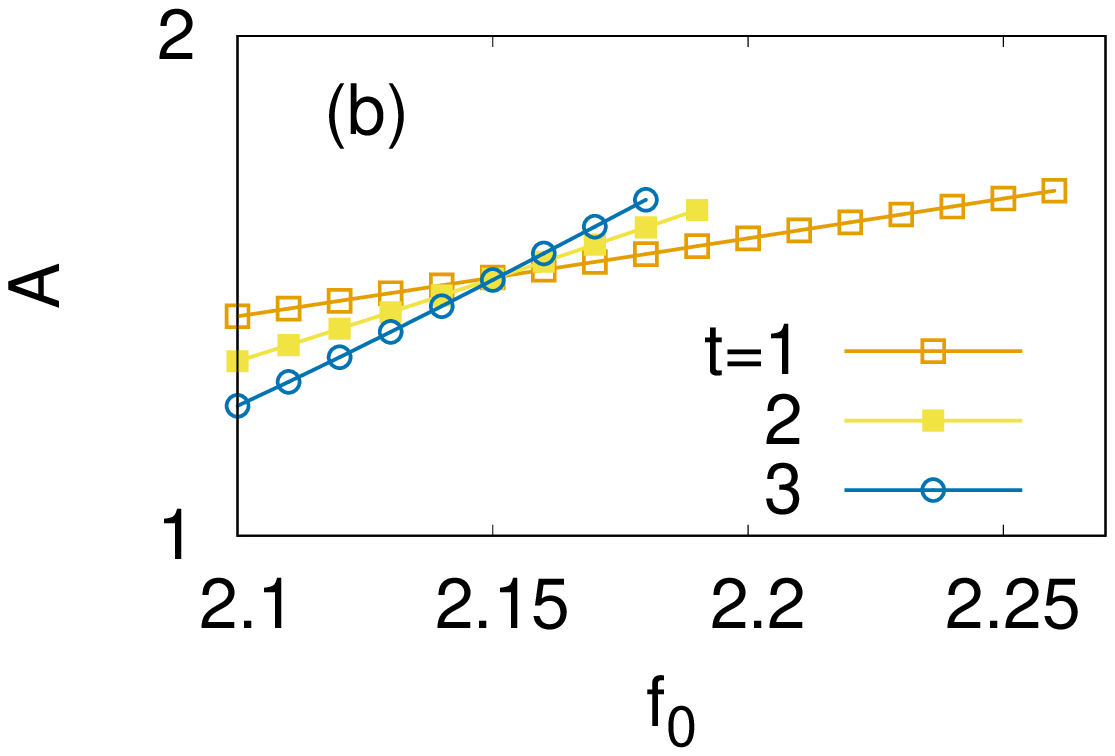}
\caption{Direct simulation of Eq.~\eqref{eq:kow} using the spectral method. The
values of the modeling parameters are the same as in the gradient descent
calculation in Fig.~\ref{fig:grad}. We have chosen $P=M=3$ in
Eq.~\eqref{eq:basis}, and the time step size for the Euler scheme is $\Delta t =
10^{-3}$.
(a) For each different $f_0$, the initial condition is set to be a uniform
distribution of $\rho_0 = 1$ and $c_0 = (f_0 / g_0) \rho_0$ with small random
perturbation in the coefficients from $E_{pm}$ to $G_{pm}$.
The perturbation grows for $f_0 \gtrsim f_0^\text{unstable} \approx 2.195$.
(b) For every $f_0$, the initial condition is a postaggregation profile,
obtained from (a) by running the system for $t=50$ at $f_0 = 2.26 >
f_0^\text{unstable}$. The aggregate dissolves at $f_0 \lesssim 2.15$,
differently from the threshold behavior in (a), which gives a clear signal of
hysteresis.
}
\label{fig:spec}
\end{figure}

For comparison, we directly simulate
Eq.~\eqref{eq:kow} with the spectral
method. Again, the basic idea is to expand the density fields in terms of the
Fourier-Bessel modes with time-dependent coefficients as follows:
\begin{subequations}
\label{eq:basis}
\begin{align}
\begin{split}
\rho^\text{trunc}(r,\theta,t) &= \rho_0 + \sum_{p=0}^{P-1} \sum_{m=1}^M
J_p(j'_{pm}r/l)\\&
\times \left[ E_{pm}(t) \cos (p\theta) + F_{pm}(t) \sin (p\theta) \right]
\end{split}
\\
\begin{split}
c^\text{trunc}(r,\theta,t) &= c_0(t) + \sum_{p=0}^{P-1} \sum_{m=1}^M
J_p(j'_{pm}r/l)\\&
\times \left[ G_{pm}(t) \cos (p\theta) + H_{pm}(t) \sin (p\theta) \right],
\end{split}
\end{align}
\end{subequations}
where the superscript on the left-hand side means that the expansion has
been truncated with certain positive integers $P$ and $M$. This representation
is mathematically equivalent to Eq.~\eqref{eq:expand}, and it is preferable in
the spectral method to work with the orthogonal basis set $\{ 1, J_p(j'_{pm}r/l)
\cos (p\theta), J_q(j'_{qn}r/l) \sin (q\theta) | p=0,1,\ldots, m=1,2,\ldots,
q=0,1,\ldots, n=1,2,\ldots\}$.
We substitute Eq.~\eqref{eq:basis} into the right-hand side of
Eq.~\eqref{eq:kow} and project the result onto each of
the orthogonal bases to find the time derivative of each coefficient. For
example, we have
\begin{equation}
\begin{split}
\frac{dG_{pm}}{dt} &\approx \Lambda_{pm}^{-1} \int_0^{2\pi} d\theta
\int_0^l dr ~r J_p (j'_{pm}r/l) \cos (p\theta)\\
&\times \left( f_0 \rho^\text{trunc} + \nu_0 \nabla^2 c^\text{trunc} - g_0
c^\text{trunc} \right),
\end{split}
\label{eq:proj}
\end{equation}
where $\Lambda_{pm} \equiv \int_0^{2\pi} d\theta
\int_0^l dr~r J_p^2(j'_{pm}r/l) \cos^2
(p\theta)$ is a normalization constant.
We use the Euler scheme to update the coefficients, which means that
Eq.~\eqref{eq:proj} is implemented as $G_{pm} (t+\Delta t) \approx G_{pm}(t) +
(dG_{pm}/dt) \Delta t$ with a small step size $\Delta t$.
An advantage of the spectral method is that it correctly preserves the
total mass $M$, and the price is repeated numerical integration
[see Eq.~\eqref{eq:proj}], which we have carried out by means of the Gaussian
quadrature~\cite{newman2013computational}.

Figure~\ref{fig:spec} shows our simulation results. When we start from a uniform
distribution with small perturbation, aggregation develops for $f_0 \gtrsim
f_0^\text{unstable} \approx 2.195$ [Fig.~\ref{fig:spec}(a)]. On the other hand,
if the simulation starts from an aggregate, then it persists even if $f_0$ goes
below $f_0^\text{unstable}$ [Fig.~\ref{fig:spec}(b)]. Such hysteresis behavior,
a characteristic of discontinuous phase transition, is fully consistent with the
gradient descent result of the Lyapunov functional.

\section{Discussion}
\label{sec:discussion}

We have observed two points from our model: First,
aggregation occurs discontinuously. It means that an aggregate
forms abruptly as the chemotactic coupling strength increases, and once it
happens, it will maintain itself even if the coupling weakens again. Another
point is that an aggregate forms in the vicinity of the boundary. This is also
plausible if we think of smaller motility near the boundary than in the
middle due to the zero-flux boundary conditions.
If the disk geometry is understood as a projected image of an infinite
cylinder, then this picture may serve as a simple model for bacterial
aggregation inside a blood vessel (see, e.g.,
Refs.~\onlinecite{sittner2017pathology,bonazzi2018intermittent}). In this
context, we can see the biomedical importance of understanding the
mechanism of aggregation, considering that aggregation increases antibiotic
resistance structurally even without altering the innate mechanism of individual
cells~\cite{stewart2001antibiotic,secor2018entropically}. Our result
implies that a bacterial aggregate will start to form near the blood vessel
wall and that it will not smoothly disappear even if the environmental condition
is restored.

The two observations--the boundary effect and discontinuity--are
actually related, and the transition behavior may change if we suppress
aggregation at the boundary. To see this, let us restrict ourselves to a
radially symmetric case ($p=0$) with $m=1$ and $2$. We have to assume $R_{01} >
0$ to consider aggregation in the middle of the disk. As above, we write an
approximate expression for $W_\text{st}$ with these two modes and check its
minimum at $(R_{01}^*, R_{02}^*)$. When $R_{01}^*$ is small near the transition
point, we can show that
\begin{equation}
R_{02}^* = \frac{\kappa z_{01}^2 \int_0^l r J_0^2(j'_{01}r/l)
J_0(j'_{02}r/l) dr}
{J_0^2(j'_{02}) l^2 [(1+\kappa \rho_0) z_{02} - \chi_0/D_0]} {R_{01}^*}^2
+ O\left( {R_{01}^*}^3 \right).
\label{eq:taylor}
\end{equation}
Because of this, the lowest-order nonvanishing term involving $R_{02}$ in
$W_\text{st}$ is effectively of an order of $R_{01}^4$ around the minimum
(Appendix~\ref{appendix:radial}). If the effect of $m=2$ is negligible, then, by
the same token, it can be argued that the next modes with
$m>2$ will also be small compared to $m=1$. The Lyapunov functional can thus be
further approximated as a cubic polynomial in $R_{01}$, as one sees in
mean-field $\varphi^3$ theory~\cite{bender2013observation}. The resulting phase
transition is continuous, and the order parameter increases linearly with a
slope $\propto \kappa^{-1}$ when the control parameter exceeds the transition
threshold. It contrasts with the classical PKS model, which
exhibits a discontinuous transition in two dimensions even under the restriction
to $p=0$~\cite{childress1981nonlinear} because its pressure function $\Pi$
has only logarithmic dependence on $\rho$~\cite{kowalczyk2005preventing}.

From a theoretical point of view, the Lyapunov-functional approach provides an
intriguing way to consider free energy for active matter. We have seen that the
functional greatly helps us understand the aggregation phenomenon of active
particles with chemotactic interaction, just as conventional free energy does in
equilibrium statistical physics. Obviously, this approach is not always
applicable because most dynamical systems have no Lyapunov functional, and even
if a Lyapunov functional exists, we have no systematic method to construct it.
Nevertheless, it is possible to derive the functional for a class of chemotactic
systems with arbitrary $\rho$ dependence of pressure as in
Appendix~\ref{appendix:lyapunov}. For such a class of systems, the dynamics is
essentially relaxational, and in this sense, they still belong to a tractable
category of nonequilibrium although they deal with active matter. Finding a
general form of models in this category will certainly be an interesting
topic, and one may well ask, for example, whether this approach can cover
recently observed effects of microswimmers such as self-organized long-ranged
flows or hydrodynamically mediated
interactions~\cite{mathijssen2018nutrient,mathijssen2018universal}.

By analogy between the Lyapunov functional and free energy, it would also be
interesting to ask the meaning of a derivative of $W_\text{st}$ with respect to
a system parameter, say, $f_0$. Figure~\ref{fig:grad}(a) then implies that the
derivative may serve as an alternative order parameter, showing similar behavior
to that of $A$ in Fig.~\ref{fig:grad}(b). One way to interpret this derivative
is to note that we can produce a similar effect on the system by modulating
either $f_0$ or $l$ [see Eq.~\eqref{eq:zpm}]. In other words, the system will
behave in qualitatively the same way as in Fig.~\ref{fig:grad}(a) when
$V=\pi l^2$ expands. Then, one could understand $-\partial
W^\text{min}_\text{st}/\partial V$ as a kind of pressure exerted on
$\partial\Omega$ because pressure and volume constitute a conjugate pair in
thermodynamics. Of course, this pressure is an ensemble-averaged value and
should not be confused with $\Pi$. Whether it can be regarded as a physical
pressure is a challenging question, which asks if the Lyapunov functional is
anything more than a mere mathematical construct. Full thermodynamic
understanding of this Lyapunov-functional approach is yet to be developed.

\section{Summary}
\label{sec:summary}

To summarize, we have investigated a two-dimensional system of chemotactically
interacting organisms which experience effective pressure as a function of their
own local density. Our main analytic tool has been the Lyapunov functional,
whose landscape can be seen most clearly through a mode expansion
[Eq.~\eqref{eq:expand} or \eqref{eq:basis}]. To couple the modes
in the simplest way, we have introduced a quadratic term in the
virial expansion of effective pressure. Our main interest is the transition
behavior to aggregation, and we have observed clear signals of a discontinuous
transition (Figs.~\ref{fig:grad} and \ref{fig:spec}).
By using a two-mode approximation, we have
provided a qualitative explanation for the transition behavior
[Figs.~\ref{fig:xy}(c) and \ref{fig:xy}(d)]. We have also discussed a special
case under radial symmetry and proposed a thermodynamically motivated order
parameter for aggregation transition, defined as a first-order derivative of
free energy.

\begin{acknowledgments}
This work was supported by Basic Science Research Program through the
National Research Foundation of Korea (NRF) funded by the Ministry of Science,
ICT and Future Planning (Grant No. NRF-2017R1A1A1A05001482).
\end{acknowledgments}

\appendix

\begin{widetext}
\section{Derivation of the Lyapunov functional}
\label{appendix:lyapunov}

Let us define $\tau \equiv t/D_0$, $\tilde{c} \equiv (\chi_0 / D_0)c$,
$\tilde{\Pi}(\rho) = \Pi(\rho) / D_0$, and $Z \equiv -\tilde{c} + \tilde{\Pi}$.
The model is expressed as
\begin{subequations}
\begin{align}
\frac{\partial\rho}{\partial \tau} &= \nabla \cdot \left( \rho \nabla
Z \right)\label{eq:kow5}\\
\frac{\chi_0}{\nu_0}\frac{\partial \tilde{c}}{\partial \tau} &=
\lambda \rho +  \nabla^2 \tilde{c} - \frac{g_0}{\nu_0}
\tilde{c}.\label{eq:kow6}
\end{align}
\end{subequations}
The total mass $M \equiv \int \rho ~dV$ is conserved. Denoting the
problem domain as $\Omega$, we impose the Neumann boundary conditions,
\begin{equation}
\left.\frac{\partial{\rho}}{\partial{r}} \right|_{\partial\Omega} =
\left.\frac{\partial{c}}{\partial{r}} \right|_{\partial\Omega} = 0.
\end{equation}
We note that the following equality holds:
\begin{eqnarray}
\frac{d}{d\tau} \int dV Z \rho &=&
\int dV \rho \frac{\partial Z}{\partial \tau} +
\int dV Z\frac{\partial \rho}{\partial \tau} \label{eq:ddt}\\
&=&
\int dV \rho \frac{\partial Z}{\partial \tau} - \int dV \rho |\nabla Z|^2
\label{eq:ddt2}
\end{eqnarray}
because the last term on the right-hand side of Eq.~\eqref{eq:ddt} can be
rewritten as follows:
\begin{eqnarray}
\int dV \nabla \cdot (Z\rho \nabla Z) - \int dV \rho |\nabla Z|^2
&=& \int dV Z \nabla \cdot (\rho \nabla Z)\label{eq:div}\\
&=& \int dV Z\frac{\partial \rho}{\partial \tau},
\end{eqnarray}
where the first term of the left-hand side of Eq.~\eqref{eq:div} is
zero due to the boundary conditions.
Plugging the explicit expression of $Z$ into the first term on the right-hand
side of Eq.~\eqref{eq:ddt2}, we obtain
\begin{equation}
\frac{d}{d\tau} \int dV Z \rho = -\int dV \rho \frac{\partial
\tilde{c}}{\partial \tau} + \int dV \rho \frac{\partial \tilde{\Pi}}{\partial
\tau} - \int dV \rho |\nabla Z|^2.
\label{eq:ddt3}
\end{equation}
Through integration by parts, this becomes
\begin{equation}
\frac{d}{d\tau} \int dV (-c') \rho = -\int dV \rho \frac{\partial
\tilde{c}}{\partial \tau} - \int dV \tilde{\Pi} \frac{\partial \rho}{\partial
\tau} - \int dV \rho |\nabla Z|^2.
\label{eq:ddt4}
\end{equation}
If we introduce $\Psi(\rho)$ such that $d\Psi / d\rho = \tilde{\Pi}$, it means
\begin{equation}
\frac{d}{d\tau} \int dV \left( \Psi - \rho c' \right)
= -\int dV \rho \frac{\partial \tilde{c}}{\partial \tau}
- \int dV \rho |\nabla Z|^2.
\label{eq:ddt5}
\end{equation}
To deal with the first term on the right-hand side of Eq.~\eqref{eq:ddt5},
we derive from the boundary conditions the following identity:
\begin{eqnarray}
0 &=& \int dV \nabla \cdot \left( \frac{\partial \tilde{c}}{\partial \tau}
\nabla \tilde{c} \right)\\
&=& \int dV \frac{\partial c'}{\partial \tau} \nabla^2 c'
+ \frac{d}{d\tau} \int dV \frac{|\nabla \tilde{c}|^2}{2}\\
&=& \int dV \frac{\partial c'}{\partial \tau} \left(
\frac{\chi_0}{\nu_0} \frac{\partial c'}{\partial \tau} - \lambda \rho +
\frac{g_0}{\nu_0} c' \right)
+ \frac{d}{d\tau} \int dV \frac{|\nabla \tilde{c}|^2}{2}\\
&=& \frac{\chi_0}{\nu_0} \int dV \left( \frac{\partial c ^{\prime}}{\partial
\tau} \right)^2
- \lambda\int dV \rho \frac{\partial c ^{\prime}}{\partial \tau}
+ \frac{g_0}{\nu_0} \frac{d}{d\tau} \int dV \frac{{\tilde{c}}^2}{2}
+ \frac{d}{d\tau} \int dV \frac{|\nabla \tilde{c}|^2}{2},
\end{eqnarray}
so that
\begin{equation}
\lambda\int dV \rho \frac{\partial c ^{\prime}}{\partial \tau}
=
\frac{\chi_0}{\nu_0} \int dV \left( \frac{\partial c ^{\prime}}{\partial
\tau} \right)^2
+ \frac{g_0}{\nu_0} \frac{d}{d\tau} \int dV \frac{{\tilde{c}}^2}{2}
+ \frac{d}{d\tau} \int dV \frac{|\nabla \tilde{c}|^2}{2}.
\label{eq:ddt6}
\end{equation}
By substituting Eq.~\eqref{eq:ddt6} into Eq.~\eqref{eq:ddt5} and rearranging the
terms, we get
\begin{equation}
-\frac{dW}{d\tau}  = \frac{\chi_0}{\nu_0} \int dV \left( \frac{\partial c
^{\prime}}{\partial \tau} \right)^2 + \lambda \int dV \rho
|\nabla Z|^2,
\end{equation}
where
\begin{equation}
W \equiv \lambda \int dV \left( \Psi - \rho \tilde{c} \right)
+\frac{g_0}{2\nu_0}
\int dV {\tilde{c}}^2 + \int dV \frac{|\nabla \tilde{c}|^2}{2}
\end{equation}
is the Lyapunov functional in the sense that $dW/dt \le 0$.

\section{Lyapunov functional for $\kappa=0$}
\label{sec:series}

Let us recall that the Bessel function satisfies the
following equation~\cite{boas2006mathematical}:
\begin{equation}
\int_0^1 x J_p(\alpha x) J_p(\beta x) dx
= - \frac{\alpha J_p(\beta) \frac{d}{dx} J_p
(\alpha) - \beta \frac{d}{dx}J_p(\beta) J_p(\alpha)}{\alpha^2 - \beta^2}
\label{eq:boas}
\end{equation}
if $\alpha \neq \beta$. If $\alpha = j'_{pm}$ and $\beta = j'_{pn}$ with $m \neq
n$, the right-hand side vanishes because $\frac{d}{dx}J_p(\alpha) =
\frac{d}{dx}J_p(\beta) = 0$. On the other hand, if $\alpha = j'_{pm}$ and
$\beta$ is an arbitrary number, then we instead have
\begin{equation}
\int_0^1 x J_p(\alpha x) J_p(\beta x) dx
= \frac{\beta \frac{d}{dx}J_p(\beta) J_p(\alpha)}{\alpha^2 - \beta^2}.
\end{equation}
In the limit of $\beta \to \alpha$, we obtain
\begin{equation}
\int_0^1 x J_p(\alpha x) J_p(\alpha x) dx
= -\frac{1}{2} J_p(\alpha) \frac{d^2}{dx^2}J_p(\alpha)
\end{equation}
for $\alpha = j'_{pm}$. We thus have the following orthogonality relation:
\begin{equation}
\int_0^l r J_p(j'_{pm}r/l) J_p(j'_{pn}r/l) ~dr
= -\frac{l^2}{2} \zeta_{pm} \delta_{mn},
\label{eq:ortho}
\end{equation}
where $\delta_{mn}$ is the Kronecker $\delta$ and
\begin{eqnarray}
\zeta_{pm} &\equiv& J_p (j'_{pm}) \frac{d^2}{dx^2} J_p (j'_{pm})\\
&=& \frac{1}{4} J_p (j'_{pm}) \left[ J_{p-2}(j'_{pm}) -2J_p(j'_{pm}) +
J_{p+2}(j'_{pm})\right]\\
&=& \frac{1}{4} J_p (j'_{pm}) \left[ \frac{2(p-1)}{j'_{pm}}J_{p-1}(j'_{pm})
+ \frac{2(p+1)}{j'_{pm}} J_{p+1}(j'_{pm}) -4 J_p(j'_{pm}) \right]\\
&=& \frac{1}{4} J_p (j'_{pm}) \left[ \frac{2(p-1)}{j'_{pm}} \frac{p}{j'_{pm}}
J_p(j'_{pm}) + \frac{2(p+1)}{j'_{pm}} \frac{p}{j'_{pm}} J_p(j'_{pm}) -4
J_p(j'_{pm}) \right]\\
&=& J_p^2(j'_{pm}) (p^2/j'^2_{pm}-1).
\end{eqnarray}
From this orthogonality relation, it follows that
\begin{eqnarray}
\frac{1}{\pi l^2} \int \rho^2 dV &=& \rho_0^2 - \sum_{m=1}^\infty \left(
\zeta_{0m} Q_{0m}^2 + \frac{1}{2} \sum_{p=1}^\infty \zeta_{pm} Q_{pm}^2 \right)
\label{eq:rho2}\\
\frac{1}{\pi l^2} \int c^2 dV &=& c_0^2 - \sum_{m=1}^\infty \left( \zeta_{0m}
R_{0m}^2 + \frac{1}{2} \sum_{p=1}^\infty \zeta_{pm} R_{pm}^2 \right)
\label{eq:c2}\\
\frac{1}{\pi l^2} \int \rho c~ dV &=& \rho_0 c_0 - \sum_{m=1}^\infty \left[
\zeta_{0m} Q_{0m} R_{0m} + \frac{1}{2} \sum_{p=1}^\infty \zeta_{pm} Q_{pm}
R_{pm} \cos p(\eta_{pm}-\phi_{pm}) \right]. \label{eq:rho_c}
\end{eqnarray}
The squared gradient term can be calculated as follows:
\begin{equation}
\int |\nabla c|^2 dV =
\int_{-\pi}^{\pi} \int_0^l \left(
\left| \frac{\partial c}{\partial r} \right|^2 + \frac{1}{r^2} \left|
\frac{\partial c}{\partial \theta} \right|^2 \right) r~dr~ d\theta.
\label{eq:target}
\end{equation}
From Eqs.~\eqref{eq:rho_expand} and \eqref{eq:c_expand}, we have
\begin{eqnarray}
\frac{\partial c}{\partial r} &=& \sum_{p=0}^\infty \sum_{m=1}^\infty
\frac{j'_{pm}}{l} \left[\left.
\frac{dJ_p(x)}{dx} \right|_{x=j'_{pm}r/l} \right] R_{pm} \cos
p(\theta-\phi_{pm})\\
&=& \sum_{p=0}^\infty \sum_{m=1}^\infty
\frac{j'_{pm}}{l} \left[
\frac{J_{p-1} (j'_{pm}r/l) - J_{p+1} (j'_{pm}r/l)}{2} \right] R_{pm} \cos
p(\theta-\phi_{pm})\\
\frac{1}{r} \frac{\partial c}{\partial \theta} &=& \sum_{pm} J_p(j'_{pm}r/l)
\left( \frac{-p}{r} \right) R_{pm} \sin p(\theta - \phi_{pm})\\
&=& -\sum_{pm} \frac{j'_{pm}}{l} \left[ \frac{J_{p-1} (j'_{pm}r/l) +
J_{p+1}(j'_{pm}r/l)}{2} \right] R_{pm} \sin p(\theta - \phi_{pm}),
\end{eqnarray}
where we have used the recurrence relations for the Bessel functions,
$J_{p-1}(x)-J_{p+1}(x)=2\frac{d}{dx}J_p(x)$ and
$J_{p-1}(x)+J_{p+1}(x)=\frac{2p}{x}J_p(x)$. We substitute these expressions
into Eq.~\eqref{eq:target} and perform the integration. We first perform the
integral over $\theta$ by using the following orthogonality:
\begin{equation}
\int_{-\pi}^{\pi} \cos p(\theta-\phi_{pm}) \cos q(\theta-\phi_{qn}) d\theta =
\left\{ \begin{array}{lcl}
0 & \text{~if~} & p \neq q\\
\pi \cos p(\phi_{pm}-\phi_{qn}) & \text{~if~} & p=q \neq 0\\
2\pi & \text{~if~} & p=q=0.
\end{array}
\right.
\end{equation}
We then proceed to the integral over $r$, for which we have to evaluate the
following:
\begin{eqnarray}
\int |\nabla c|^2 dV &=& \sum_{mn} j'_{0m} j'_{0n} 2\pi R_{0m}
R_{0n} T_{mn} \nonumber\\
&+& \sum_{p=1}^\infty \sum_{mn} \frac{j'_{pm} j'_{pn}}{2} \pi R_{pm} R_{pn} \cos
p(\phi_{pm}-\phi_{pn}) S_{p,mn},
\label{eq:target2}
\end{eqnarray}
where
\begin{eqnarray}
T_{mn} &\equiv& l^{-2} \int_0^l dr ~r J_1 (j'_{0m}r/l) J_1 (j'_{0n}r/l)\\
S_{p,mn} &\equiv&
l^{-2} \int_0^l dr~r \left[ J_{p-1}(j'_{pm}r/l) J_{p-1}(j'_{pn}r/l)
+ J_{p+1}(j'_{pm}r/l) J_{p+1}(j'_{pn}r/l) \right].
\end{eqnarray}
The first integral is straightforward because $j'_{0m}$'s are the zeros of
$J_1(x) = -\frac{d}{dx}J_0(x)$. The standard orthogonality condition for the
Bessel function gives $T_{mn} = \frac{1}{2} J_0^2(j'_{0m}) \delta_{mn}$.
For $S_{p,mn}$, we use Eq.~\eqref{eq:boas} together with
the following recurrence relations:
\begin{eqnarray}
\frac{d}{dx}J_{p-1}(x) &=& \frac{p-1}{x}J_{p-1}(x) - J_p(x)\\
\frac{d}{dx}J_{p+1}(x) &=& -\frac{p+1}{x} J_{p+1}(x) + J_p(x)\\
J_{p+1}(x) &=& \frac{2p}{x} J_p(x) - J_{p-1}(x)
\end{eqnarray}
to show
\begin{eqnarray}
&&\int_0^1 dx~x J_{p-1}(\alpha x) J_{p-1}(\beta x)
+ \int_0^1 dx~x J_{p+1}(\alpha x) J_{p+1}(\beta x)\\
&=& \frac{2\alpha J_{p-1}(\beta) J_p(\alpha)}{\alpha^2-\beta^2}
- \frac{2\beta J_{p-1}(\alpha) J_p(\beta)}{\alpha^2-\beta^2}
- \frac{2p J_p(\alpha) J_p(\beta)}{\alpha\beta}.
\label{eq:mathematica}
\end{eqnarray}
This is true for arbitrary $\alpha$ and $\beta$, so it implies that
\begin{equation}
S_{p,mn} = \frac{2j'_{pm} J_{p-1}(j'_{pn}) J_p(j'_{pm})}{j'^2_{pm}-j'^2_{pn}}
- \frac{2j'_{pn} J_{p-1}(j'_{pm}) J_p(j'_{pn})}{j'^2_{pm}-j'^2_{pn}}
- \frac{2p J_p(j'_{pm}) J_p(j'_{pn})}{j'_{pm} j'_{pn}}
\label{eq:mathematica2}
\end{equation}
for $m \neq n$. By definition, we also have $\frac{d}{dx}J_p(j'_{pm}) =
\frac{d}{dx}J_p(j'_{pn}) = 0$, which means that $J_{p-1} (j'_{pm}) =
\frac{p}{j'_{pm}} J_p(j'_{pm})$ and $J_{p-1} (j'_{pn}) = \frac{p}{j'_{pn}}
J_p(j'_{pn})$ because $\frac{d}{dx} J_p(x) = -\frac{p}{x} J_p(x) + J_{p-1}(x)$.
If we plug them into Eq.~\eqref{eq:mathematica2}, then we find that
$S_{p,mn}$ identically vanishes for $m \neq n$.
On the other hand, by taking the limit of $\beta \to \alpha$ in
Eq.~\eqref{eq:mathematica}, we find
\begin{eqnarray}
&&\int_0^1 dx~x J_{p-1}(\alpha x) J_{p-1}(\alpha x)
+ \int_0^1 dx~x J_{p+1}(\alpha x) J_{p+1}(\alpha x)\nonumber\\
&=& \frac{1}{\alpha^2} \left\{ (\alpha^2-2p) J_p^2(\alpha) + \alpha
J_{p-1}(\alpha) \left[ 2J_p(\alpha) - \alpha J_{p+1}(\alpha) \right] \right\}.
\end{eqnarray}
If $\alpha = j'_{pm}$, then this expression reduces to
$\left( 1-\frac{p^2}{\alpha^2} \right) J_p^2(\alpha)$
because $J_{p+1}(\alpha) = J_{p-1}(\alpha) =
\frac{p}{\alpha} J_p(\alpha)$ from $\frac{d}{dx}J_p(x) = \frac{p}{x} J_p(x) -
J_{p+1}(x) = -\frac{p}{x} J_p(x) + J_{p-1}(x)$. We summarize the above result as
\begin{equation}
S_{p,mn} = \left(1-\frac{p^2}{j'^2_{pm}} \right) J_p^2(j'_{pm}) \delta_{mn}.
\end{equation}
Consequently, Eq.~\eqref{eq:target2} yields the following:
\begin{equation}
\int |\nabla c|^2 dV
= \sum_{m=1}^\infty \pi R_{0m}^2 j'^2_{0m} J_0^2(j'_{0m})
+ \frac{1}{2} \sum_{p=1}^\infty \sum_{m=1}^\infty \pi R^2_{pm} (j'^2_{pm}-p^2)
J_p^2(j'_{pm}).
\label{eq:grad}
\end{equation}
Furthermore, Eq.~\eqref{eq:kow4} implies that
\begin{subequations}
\label{eq:stationary}
\begin{align}
\rho_0 &= \frac{g_0}{f_0} c_0\\
Q_{pm} &= \left( \frac{\nu_0}{f_0} \frac{j'^2_{pm}}{l^2} + \frac{g_0}{f_0}
\right) R_{pm} = z_{pm} R_{pm}\\
\eta_{pm} &= \phi_{pm}
\end{align}
\end{subequations}
in the stationary state at $t \to \infty$. By using Eqs.~\eqref{eq:rho2},
\eqref{eq:c2}, \eqref{eq:rho_c}, \eqref{eq:grad}, and \eqref{eq:stationary}, we
obtain Eq.~\eqref{eq:quad} of the main text.

\section{Linear stability analysis}
\label{appendix:linear}

Let us assume that the homogeneous solution is perturbed by a small amount
$\epsilon$ as follows:
\begin{subequations}
\begin{align}
\rho(r,t) &= \rho_0 +\epsilon_{\rho}\\
c(r,t) &= \frac{f_0}{g_0}\rho_0 +\epsilon_c.
\end{align}
\end{subequations}
Substituting these expressions into the governing equation
[Eq.~\eqref{eq:kow}] and expanding the result
to the first order of $\epsilon$'s, we see
\begin{equation}
\frac{\partial}{\partial{t}}
\begin{pmatrix}\epsilon_\rho\\
\epsilon_c\\ \end{pmatrix} =
\begin{pmatrix}
0 & 0 \\
f_0 & -g_0 \\
\end{pmatrix}
\begin{pmatrix}\epsilon_\rho\\
\epsilon_c\\ \end{pmatrix}
+ \begin{pmatrix}
D_0 \rho_0 + D_1 \rho^2_0 & -\chi_0 \rho_0 \\
0 & \nu_0 \\
\end{pmatrix}
\nabla^2
\begin{pmatrix}\epsilon_\rho\\
\epsilon_c\\ \end{pmatrix}.
\label{eq:linear}
\end{equation}
The solution of this linear equation is assumed to take a form of
$J_p(kr)e^{ip\theta} e^{\eta t}$,
where $J_p$ means the Bessel function of order $p$,
and $\eta$ is the growth rate of the perturbation. The boundary conditions
dictate that $(\partial/\partial r) J_p(kl) = 0$. The
lowest-wavenumber mode is given by $kl \approx 1.841$ for $p=1$.
Noting that $(\nabla^2 + k^2) [J_p (kr) e^{ip\theta}] =
0$, Eq.~\eqref{eq:linear} reduces to the following equation for $\eta$:
\begin{equation}
\eta^2 +\left[k^2(D_{0}\rho_0 + D_{1}\rho^2_0) + g_0
+k^{2}\nu_{0}\right] + k^2(D_{0}\rho_0 +
D_{1}\rho^2_0)(g_0 +k^{2}\nu_{0}).
\end{equation}
For $\eta$ to be negative, the following inequality must be met:
\begin{equation}
k^2 > \left(1+\frac{D_1}{D_0}\rho_0 \right)^{-1} \frac{f_0
\chi_0}{\nu_0 D_0} - \frac{g_0}{\nu_0},
\end{equation}
which is the stability condition.

\section{Two-mode approximation under radial symmetry}
\label{appendix:radial}

When we take $\rho \approx \rho_0 + Q_{01} J_0 (j'_{01} r/l) + Q_{02} J_0
(j'_{02} r/l)$, the corresponding Lyapunov functional becomes
\begin{eqnarray}
\frac{W_\text{st}}{V} &=& \frac{W_\text{hom}}{V} +
\frac{\lambda}{2} \sum_{m=1}^2 J_0^2(j'_{0m}) z_{0m} \left[
(1+\kappa \rho_0) z_{0m} - \frac{\chi_0}{D_0} \right] R_{0m}^2\nonumber\\
&+& \frac{\kappa \lambda}{3l^2} \left( z_{01}^3 R_{01}^3 I_{111} + 3z_{01}^2
z_{02} R_{01}^2 R_{02} I_{112} + 3z_{01}z_{02}^2 R_{01}R_{02}^2 I_{122} +
z_{02}^3 R_{02}^3 I_{222} \right),
\label{eq:radialw}
\end{eqnarray}
where $I_{ijk} \equiv \int_0^l r J_0(j'_{0i}r/l) J_0(j'_{0j}r/l)
J_0(j'_{0k}r/l) dr$.
Our assumption is that $(1+\kappa \rho_0) z_{01} - \frac{\chi_0}{D_0} < 0$
and $(1+\kappa \rho_0) z_{02} - \frac{\chi_0}{D_0} > 0$ so that the homogeneous
solution is instabilized by the first mode.
We differentiate the above expression with respect to
$R_{02}$ to find a minimum:
\begin{equation}
0 = \left.\lambda J_0^2(j'_{02}) z_{02} R_{02} \left[ (1+\kappa\rho_0) z_{02} -
\frac{\chi_0}{D_0} \right] + \frac{\kappa \lambda}{3l^2} (3z_{02}^3 R_{02}^2
I_{222} + 3z_{01}^2 z_{02} R_{01}^2 I_{112} + 6z_{01} z_{02}^2 R_{01} R_{02}
I_{122}) \right|_{(R_{01}^*, R_{02}^*)}.
\end{equation}
After solving this equation for $R_{02}^*$, we expand the solution as a
polynomial of $R_{01}^*$, and the result is obtained as follows:
\begin{equation}
R_{02}^* \approx \frac{\kappa z_{01}^2 I_{112}}{J_0^2(j'_{02}) l^2
[(1+\kappa\rho_0)z_{02} - \chi_0/D_0]} {R_{01}^*}^2.
\end{equation}
We note that Eq.~\eqref{eq:radialw} does not contain $R_{02}$ and
$R_{01}R_{02}$ due to the orthogonality between normal modes.
Therefore, when $R_{02} \sim O\left({R_{01}}^2 \right)$ around the minimum,
the lowest contribution from $R_{02}$ is of $O\left(R_{01}^4 \right)$.

\end{widetext}


\begin{thebibliography}{40}%
\makeatletter
\providecommand \@ifxundefined [1]{%
 \@ifx{#1\undefined}
}%
\providecommand \@ifnum [1]{%
 \ifnum #1\expandafter \@firstoftwo
 \else \expandafter \@secondoftwo
 \fi
}%
\providecommand \@ifx [1]{%
 \ifx #1\expandafter \@firstoftwo
 \else \expandafter \@secondoftwo
 \fi
}%
\providecommand \natexlab [1]{#1}%
\providecommand \enquote  [1]{``#1''}%
\providecommand \bibnamefont  [1]{#1}%
\providecommand \bibfnamefont [1]{#1}%
\providecommand \citenamefont [1]{#1}%
\providecommand \href@noop [0]{\@secondoftwo}%
\providecommand \href [0]{\begingroup \@sanitize@url \@href}%
\providecommand \@href[1]{\@@startlink{#1}\@@href}%
\providecommand \@@href[1]{\endgroup#1\@@endlink}%
\providecommand \@sanitize@url [0]{\catcode `\\12\catcode `\$12\catcode
  `\&12\catcode `\#12\catcode `\^12\catcode `\_12\catcode `\%12\relax}%
\providecommand \@@startlink[1]{}%
\providecommand \@@endlink[0]{}%
\providecommand \url  [0]{\begingroup\@sanitize@url \@url }%
\providecommand \@url [1]{\endgroup\@href {#1}{\urlprefix }}%
\providecommand \urlprefix  [0]{URL }%
\providecommand \Eprint [0]{\href }%
\providecommand \doibase [0]{https://doi.org/}%
\providecommand \selectlanguage [0]{\@gobble}%
\providecommand \bibinfo  [0]{\@secondoftwo}%
\providecommand \bibfield  [0]{\@secondoftwo}%
\providecommand \translation [1]{[#1]}%
\providecommand \BibitemOpen [0]{}%
\providecommand \bibitemStop [0]{}%
\providecommand \bibitemNoStop [0]{.\EOS\space}%
\providecommand \EOS [0]{\spacefactor3000\relax}%
\providecommand \BibitemShut  [1]{\csname bibitem#1\endcsname}%
\let\auto@bib@innerbib\@empty
\bibitem [{\citenamefont {Shapiro}(1988)}]{shapiro1988bacteria}%
  \BibitemOpen
  \bibfield  {author} {\bibinfo {author} {\bibfnamefont {J.~A.}\ \bibnamefont
  {Shapiro}},\ }\href@noop {} {\bibfield  {journal} {\bibinfo  {journal} {Sci.
  Am.}\ }\textbf {\bibinfo {volume} {258}},\ \bibinfo {pages} {82} (\bibinfo
  {year} {1988})}\BibitemShut {NoStop}%
\bibitem [{\citenamefont {Ma}\ and\ \citenamefont
  {Eaton}(1992)}]{ma1992multicellular}%
  \BibitemOpen
  \bibfield  {author} {\bibinfo {author} {\bibfnamefont {M.}~\bibnamefont
  {Ma}}\ and\ \bibinfo {author} {\bibfnamefont {J.~W.}\ \bibnamefont {Eaton}},\
  }\href@noop {} {\bibfield  {journal} {\bibinfo  {journal} {Proc. Natl. Acad.
  Sci. U.S.A.}\ }\textbf {\bibinfo {volume} {89}},\ \bibinfo {pages} {7924}
  (\bibinfo {year} {1992})}\BibitemShut {NoStop}%
\bibitem [{\citenamefont {Shapiro}(1995)}]{shapiro1995significances}%
  \BibitemOpen
  \bibfield  {author} {\bibinfo {author} {\bibfnamefont {J.~A.}\ \bibnamefont
  {Shapiro}},\ }\href@noop {} {\bibfield  {journal} {\bibinfo  {journal}
  {BioEssays}\ }\textbf {\bibinfo {volume} {17}},\ \bibinfo {pages} {597}
  (\bibinfo {year} {1995})}\BibitemShut {NoStop}%
\bibitem [{\citenamefont {Crespi}(2001)}]{crespi2001evolution}%
  \BibitemOpen
  \bibfield  {author} {\bibinfo {author} {\bibfnamefont {B.~J.}\ \bibnamefont
  {Crespi}},\ }\href@noop {} {\bibfield  {journal} {\bibinfo  {journal} {Trends
  Ecol. Evol.}\ }\textbf {\bibinfo {volume} {16}},\ \bibinfo {pages} {178}
  (\bibinfo {year} {2001})}\BibitemShut {NoStop}%
\bibitem [{\citenamefont {Detrain}\ and\ \citenamefont
  {Deneubourg}(2006)}]{detrain2006self}%
  \BibitemOpen
  \bibfield  {author} {\bibinfo {author} {\bibfnamefont {C.}~\bibnamefont
  {Detrain}}\ and\ \bibinfo {author} {\bibfnamefont {J.-L.}\ \bibnamefont
  {Deneubourg}},\ }\href@noop {} {\bibfield  {journal} {\bibinfo  {journal}
  {Phys. Life Rev.}\ }\textbf {\bibinfo {volume} {3}},\ \bibinfo {pages} {162}
  (\bibinfo {year} {2006})}\BibitemShut {NoStop}%
\bibitem [{\citenamefont {Stewart}\ and\ \citenamefont
  {Costerton}(2001)}]{stewart2001antibiotic}%
  \BibitemOpen
  \bibfield  {author} {\bibinfo {author} {\bibfnamefont {P.~S.}\ \bibnamefont
  {Stewart}}\ and\ \bibinfo {author} {\bibfnamefont {J.~W.}\ \bibnamefont
  {Costerton}},\ }\href@noop {} {\bibfield  {journal} {\bibinfo  {journal}
  {Lancet}\ }\textbf {\bibinfo {volume} {358}},\ \bibinfo {pages} {135}
  (\bibinfo {year} {2001})}\BibitemShut {NoStop}%
\bibitem [{\citenamefont {Secor}\ \emph {et~al.}(2018)\citenamefont {Secor},
  \citenamefont {Michaels}, \citenamefont {Ratjen}, \citenamefont {Jennings},\
  and\ \citenamefont {Singh}}]{secor2018entropically}%
  \BibitemOpen
  \bibfield  {author} {\bibinfo {author} {\bibfnamefont {P.~R.}\ \bibnamefont
  {Secor}}, \bibinfo {author} {\bibfnamefont {L.~A.}\ \bibnamefont {Michaels}},
  \bibinfo {author} {\bibfnamefont {A.}~\bibnamefont {Ratjen}}, \bibinfo
  {author} {\bibfnamefont {L.~K.}\ \bibnamefont {Jennings}},\ and\ \bibinfo
  {author} {\bibfnamefont {P.~K.}\ \bibnamefont {Singh}},\ }\href@noop {}
  {\bibfield  {journal} {\bibinfo  {journal} {Proc. Natl. Acad. Sci. U.S.A.}\
  }\textbf {\bibinfo {volume} {115}},\ \bibinfo {pages} {10780} (\bibinfo
  {year} {2018})}\BibitemShut {NoStop}%
\bibitem [{\citenamefont {Patlak}(1953)}]{patlak1953random}%
  \BibitemOpen
  \bibfield  {author} {\bibinfo {author} {\bibfnamefont {C.~S.}\ \bibnamefont
  {Patlak}},\ }\href@noop {} {\bibfield  {journal} {\bibinfo  {journal} {Bull.
  Math. Biophys.}\ }\textbf {\bibinfo {volume} {15}},\ \bibinfo {pages} {311}
  (\bibinfo {year} {1953})}\BibitemShut {NoStop}%
\bibitem [{\citenamefont {Keller}\ and\ \citenamefont
  {Segel}(1970)}]{keller1970initiation}%
  \BibitemOpen
  \bibfield  {author} {\bibinfo {author} {\bibfnamefont {E.~F.}\ \bibnamefont
  {Keller}}\ and\ \bibinfo {author} {\bibfnamefont {L.~A.}\ \bibnamefont
  {Segel}},\ }\href@noop {} {\bibfield  {journal} {\bibinfo  {journal} {J.
  Theor. Biol.}\ }\textbf {\bibinfo {volume} {26}},\ \bibinfo {pages} {399}
  (\bibinfo {year} {1970})}\BibitemShut {NoStop}%
\bibitem [{\citenamefont {Nanjundiah}(1973)}]{nanjundiah1973chemotaxis}%
  \BibitemOpen
  \bibfield  {author} {\bibinfo {author} {\bibfnamefont {V.}~\bibnamefont
  {Nanjundiah}},\ }\href@noop {} {\bibfield  {journal} {\bibinfo  {journal} {J.
  Theor. Biol.}\ }\textbf {\bibinfo {volume} {42}},\ \bibinfo {pages} {63}
  (\bibinfo {year} {1973})}\BibitemShut {NoStop}%
\bibitem [{\citenamefont {Childress}\ and\ \citenamefont
  {Percus}(1981)}]{childress1981nonlinear}%
  \BibitemOpen
  \bibfield  {author} {\bibinfo {author} {\bibfnamefont {S.}~\bibnamefont
  {Childress}}\ and\ \bibinfo {author} {\bibfnamefont {J.~K.}\ \bibnamefont
  {Percus}},\ }\href@noop {} {\bibfield  {journal} {\bibinfo  {journal} {Math.
  Biosci.}\ }\textbf {\bibinfo {volume} {56}},\ \bibinfo {pages} {217}
  (\bibinfo {year} {1981})}\BibitemShut {NoStop}%
\bibitem [{\citenamefont {Stevens}(2000)}]{stevens2000derivation}%
  \BibitemOpen
  \bibfield  {author} {\bibinfo {author} {\bibfnamefont {A.}~\bibnamefont
  {Stevens}},\ }\href@noop {} {\bibfield  {journal} {\bibinfo  {journal} {SIAM
  J. Appl. Math.}\ }\textbf {\bibinfo {volume} {61}},\ \bibinfo {pages} {183}
  (\bibinfo {year} {2000})}\BibitemShut {NoStop}%
\bibitem [{\citenamefont {Hillen}\ and\ \citenamefont
  {Painter}(2009)}]{hillen2009user}%
  \BibitemOpen
  \bibfield  {author} {\bibinfo {author} {\bibfnamefont {T.}~\bibnamefont
  {Hillen}}\ and\ \bibinfo {author} {\bibfnamefont {K.~J.}\ \bibnamefont
  {Painter}},\ }\href@noop {} {\bibfield  {journal} {\bibinfo  {journal} {J.
  Math. Biol.}\ }\textbf {\bibinfo {volume} {58}},\ \bibinfo {pages} {183}
  (\bibinfo {year} {2009})}\BibitemShut {NoStop}%
\bibitem [{\citenamefont {Gamba}\ \emph {et~al.}(2003)\citenamefont {Gamba},
  \citenamefont {Ambrosi}, \citenamefont {Coniglio}, \citenamefont {de~Candia},
  \citenamefont {Di~Talia}, \citenamefont {Giraudo}, \citenamefont {Serini},
  \citenamefont {Preziosi},\ and\ \citenamefont
  {Bussolino}}]{gamba2003percolation}%
  \BibitemOpen
  \bibfield  {author} {\bibinfo {author} {\bibfnamefont {A.}~\bibnamefont
  {Gamba}}, \bibinfo {author} {\bibfnamefont {D.}~\bibnamefont {Ambrosi}},
  \bibinfo {author} {\bibfnamefont {A.}~\bibnamefont {Coniglio}}, \bibinfo
  {author} {\bibfnamefont {A.}~\bibnamefont {de~Candia}}, \bibinfo {author}
  {\bibfnamefont {S.}~\bibnamefont {Di~Talia}}, \bibinfo {author}
  {\bibfnamefont {E.}~\bibnamefont {Giraudo}}, \bibinfo {author} {\bibfnamefont
  {G.}~\bibnamefont {Serini}}, \bibinfo {author} {\bibfnamefont
  {L.}~\bibnamefont {Preziosi}},\ and\ \bibinfo {author} {\bibfnamefont
  {F.}~\bibnamefont {Bussolino}},\ }\href@noop {} {\bibfield  {journal}
  {\bibinfo  {journal} {Phys. Rev. Lett.}\ }\textbf {\bibinfo {volume} {90}},\
  \bibinfo {pages} {118101} (\bibinfo {year} {2003})}\BibitemShut {NoStop}%
\bibitem [{\citenamefont {Kowalczyk}(2005)}]{kowalczyk2005preventing}%
  \BibitemOpen
  \bibfield  {author} {\bibinfo {author} {\bibfnamefont {R.}~\bibnamefont
  {Kowalczyk}},\ }\href@noop {} {\bibfield  {journal} {\bibinfo  {journal} {J.
  Math. Anal. Appl.}\ }\textbf {\bibinfo {volume} {305}},\ \bibinfo {pages}
  {566} (\bibinfo {year} {2005})}\BibitemShut {NoStop}%
\bibitem [{\citenamefont {Kowalczyk}\ and\ \citenamefont
  {Szyma\'{n}ska}(2008)}]{kowalczyk2008global}%
  \BibitemOpen
  \bibfield  {author} {\bibinfo {author} {\bibfnamefont {R.}~\bibnamefont
  {Kowalczyk}}\ and\ \bibinfo {author} {\bibfnamefont {Z.}~\bibnamefont
  {Szyma\'{n}ska}},\ }\href@noop {} {\bibfield  {journal} {\bibinfo  {journal}
  {J. Math. Anal. Appl.}\ }\textbf {\bibinfo {volume} {343}},\ \bibinfo {pages}
  {379} (\bibinfo {year} {2008})}\BibitemShut {NoStop}%
\bibitem [{\citenamefont {Sinhuber}\ and\ \citenamefont
  {Ouellette}(2017)}]{sinhuber2017phase}%
  \BibitemOpen
  \bibfield  {author} {\bibinfo {author} {\bibfnamefont {M.}~\bibnamefont
  {Sinhuber}}\ and\ \bibinfo {author} {\bibfnamefont {N.~T.}\ \bibnamefont
  {Ouellette}},\ }\href@noop {} {\bibfield  {journal} {\bibinfo  {journal}
  {Phys. Rev. Lett.}\ }\textbf {\bibinfo {volume} {119}},\ \bibinfo {pages}
  {178003} (\bibinfo {year} {2017})}\BibitemShut {NoStop}%
\bibitem [{\citenamefont {Wittkowski}\ \emph {et~al.}(2014)\citenamefont
  {Wittkowski}, \citenamefont {Tiribocchi}, \citenamefont {Stenhammar},
  \citenamefont {Allen}, \citenamefont {Marenduzzo},\ and\ \citenamefont
  {Cates}}]{wittkowski2014scalar}%
  \BibitemOpen
  \bibfield  {author} {\bibinfo {author} {\bibfnamefont {R.}~\bibnamefont
  {Wittkowski}}, \bibinfo {author} {\bibfnamefont {A.}~\bibnamefont
  {Tiribocchi}}, \bibinfo {author} {\bibfnamefont {J.}~\bibnamefont
  {Stenhammar}}, \bibinfo {author} {\bibfnamefont {R.~J.}\ \bibnamefont
  {Allen}}, \bibinfo {author} {\bibfnamefont {D.}~\bibnamefont {Marenduzzo}},\
  and\ \bibinfo {author} {\bibfnamefont {M.~E.}\ \bibnamefont {Cates}},\
  }\href@noop {} {\bibfield  {journal} {\bibinfo  {journal} {Nat. Commun.}\
  }\textbf {\bibinfo {volume} {5}},\ \bibinfo {pages} {4351} (\bibinfo {year}
  {2014})}\BibitemShut {NoStop}%
\bibitem [{\citenamefont {Tiribocchi}\ \emph {et~al.}(2015)\citenamefont
  {Tiribocchi}, \citenamefont {Wittkowski}, \citenamefont {Marenduzzo},\ and\
  \citenamefont {Cates}}]{tiribocchi2015active}%
  \BibitemOpen
  \bibfield  {author} {\bibinfo {author} {\bibfnamefont {A.}~\bibnamefont
  {Tiribocchi}}, \bibinfo {author} {\bibfnamefont {R.}~\bibnamefont
  {Wittkowski}}, \bibinfo {author} {\bibfnamefont {D.}~\bibnamefont
  {Marenduzzo}},\ and\ \bibinfo {author} {\bibfnamefont {M.~E.}\ \bibnamefont
  {Cates}},\ }\href@noop {} {\bibfield  {journal} {\bibinfo  {journal} {Phys.
  Rev. Lett.}\ }\textbf {\bibinfo {volume} {115}},\ \bibinfo {pages} {188302}
  (\bibinfo {year} {2015})}\BibitemShut {NoStop}%
\bibitem [{\citenamefont {Takatori}\ \emph {et~al.}(2014)\citenamefont
  {Takatori}, \citenamefont {Yan},\ and\ \citenamefont
  {Brady}}]{takatori2014swim}%
  \BibitemOpen
  \bibfield  {author} {\bibinfo {author} {\bibfnamefont {S.~C.}\ \bibnamefont
  {Takatori}}, \bibinfo {author} {\bibfnamefont {W.}~\bibnamefont {Yan}},\ and\
  \bibinfo {author} {\bibfnamefont {J.~F.}\ \bibnamefont {Brady}},\ }\href@noop
  {} {\bibfield  {journal} {\bibinfo  {journal} {Phys. Rev. Lett.}\ }\textbf
  {\bibinfo {volume} {113}},\ \bibinfo {pages} {028103} (\bibinfo {year}
  {2014})}\BibitemShut {NoStop}%
\bibitem [{\citenamefont {Takatori}\ and\ \citenamefont
  {Brady}(2015)}]{takatori2015towards}%
  \BibitemOpen
  \bibfield  {author} {\bibinfo {author} {\bibfnamefont {S.~C.}\ \bibnamefont
  {Takatori}}\ and\ \bibinfo {author} {\bibfnamefont {J.~F.}\ \bibnamefont
  {Brady}},\ }\href@noop {} {\bibfield  {journal} {\bibinfo  {journal} {Phys.
  Rev. E}\ }\textbf {\bibinfo {volume} {91}},\ \bibinfo {pages} {032117}
  (\bibinfo {year} {2015})}\BibitemShut {NoStop}%
\bibitem [{\citenamefont {Horstmann}(2001)}]{horstmann2001lyapunov}%
  \BibitemOpen
  \bibfield  {author} {\bibinfo {author} {\bibfnamefont {D.}~\bibnamefont
  {Horstmann}},\ }\href@noop {} {\bibfield  {journal} {\bibinfo  {journal}
  {Colloq. Math.}\ }\textbf {\bibinfo {volume} {87}},\ \bibinfo {pages} {113}
  (\bibinfo {year} {2001})}\BibitemShut {NoStop}%
\bibitem [{\citenamefont {Biler}(1998)}]{biler1998local}%
  \BibitemOpen
  \bibfield  {author} {\bibinfo {author} {\bibfnamefont {P.}~\bibnamefont
  {Biler}},\ }\href@noop {} {\bibfield  {journal} {\bibinfo  {journal} {Adv.
  Math. Sci. Appl.}\ }\textbf {\bibinfo {volume} {8}},\ \bibinfo {pages} {715}
  (\bibinfo {year} {1998})}\BibitemShut {NoStop}%
\bibitem [{\citenamefont {Calvez}\ and\ \citenamefont
  {Corrias}(2008)}]{calvez2008parabolic}%
  \BibitemOpen
  \bibfield  {author} {\bibinfo {author} {\bibfnamefont {V.}~\bibnamefont
  {Calvez}}\ and\ \bibinfo {author} {\bibfnamefont {L.}~\bibnamefont
  {Corrias}},\ }\href@noop {} {\bibfield  {journal} {\bibinfo  {journal}
  {Commun. Math. Sci.}\ }\textbf {\bibinfo {volume} {6}},\ \bibinfo {pages}
  {417} (\bibinfo {year} {2008})}\BibitemShut {NoStop}%
\bibitem [{\citenamefont {Fatkullin}(2013)}]{fatkullin2013study}%
  \BibitemOpen
  \bibfield  {author} {\bibinfo {author} {\bibfnamefont {I.}~\bibnamefont
  {Fatkullin}},\ }\href@noop {} {\bibfield  {journal} {\bibinfo  {journal}
  {Nonlinearity}\ }\textbf {\bibinfo {volume} {26}},\ \bibinfo {pages} {81}
  (\bibinfo {year} {2013})}\BibitemShut {NoStop}%
\bibitem [{\citenamefont {Petrosyan}\ and\ \citenamefont
  {Hu}(2014)}]{petrosyan2014nonequilibrium}%
  \BibitemOpen
  \bibfield  {author} {\bibinfo {author} {\bibfnamefont {K.~G.}\ \bibnamefont
  {Petrosyan}}\ and\ \bibinfo {author} {\bibfnamefont {C.-K.}\ \bibnamefont
  {Hu}},\ }\href@noop {} {\bibfield  {journal} {\bibinfo  {journal} {Phys. Rev.
  E}\ }\textbf {\bibinfo {volume} {89}},\ \bibinfo {pages} {042132} (\bibinfo
  {year} {2014})}\BibitemShut {NoStop}%
\bibitem [{\citenamefont {Baek}\ and\ \citenamefont
  {Kim}(2017)}]{baek2017free}%
  \BibitemOpen
  \bibfield  {author} {\bibinfo {author} {\bibfnamefont {S.~K.}\ \bibnamefont
  {Baek}}\ and\ \bibinfo {author} {\bibfnamefont {B.~J.}\ \bibnamefont {Kim}},\
  }\href@noop {} {\bibfield  {journal} {\bibinfo  {journal} {Sci. Rep.}\
  }\textbf {\bibinfo {volume} {7}},\ \bibinfo {pages} {8909} (\bibinfo {year}
  {2017})}\BibitemShut {NoStop}%
\bibitem [{\citenamefont {Cross}\ and\ \citenamefont
  {Hohenberg}(1993)}]{cross1993pattern}%
  \BibitemOpen
  \bibfield  {author} {\bibinfo {author} {\bibfnamefont {M.~C.}\ \bibnamefont
  {Cross}}\ and\ \bibinfo {author} {\bibfnamefont {P.~C.}\ \bibnamefont
  {Hohenberg}},\ }\href@noop {} {\bibfield  {journal} {\bibinfo  {journal}
  {Rev. Mod. Phys.}\ }\textbf {\bibinfo {volume} {65}},\ \bibinfo {pages} {851}
  (\bibinfo {year} {1993})}\BibitemShut {NoStop}%
\bibitem [{\citenamefont {Fanelli}\ \emph {et~al.}(2013)\citenamefont
  {Fanelli}, \citenamefont {Cianci},\ and\ \citenamefont
  {Di~Patti}}]{fanelli2013turing}%
  \BibitemOpen
  \bibfield  {author} {\bibinfo {author} {\bibfnamefont {D.}~\bibnamefont
  {Fanelli}}, \bibinfo {author} {\bibfnamefont {C.}~\bibnamefont {Cianci}},\
  and\ \bibinfo {author} {\bibfnamefont {F.}~\bibnamefont {Di~Patti}},\
  }\href@noop {} {\bibfield  {journal} {\bibinfo  {journal} {Eur. Phys. J. B}\
  }\textbf {\bibinfo {volume} {86}},\ \bibinfo {pages} {142} (\bibinfo {year}
  {2013})}\BibitemShut {NoStop}%
\bibitem [{\citenamefont {Madzvamuse}\ \emph {et~al.}(2015)\citenamefont
  {Madzvamuse}, \citenamefont {Ndakwo},\ and\ \citenamefont
  {Barreira}}]{madzvamuse2015cross}%
  \BibitemOpen
  \bibfield  {author} {\bibinfo {author} {\bibfnamefont {A.}~\bibnamefont
  {Madzvamuse}}, \bibinfo {author} {\bibfnamefont {H.~S.}\ \bibnamefont
  {Ndakwo}},\ and\ \bibinfo {author} {\bibfnamefont {R.}~\bibnamefont
  {Barreira}},\ }\href@noop {} {\bibfield  {journal} {\bibinfo  {journal} {J.
  Math. Biol.}\ }\textbf {\bibinfo {volume} {70}},\ \bibinfo {pages} {709}
  (\bibinfo {year} {2015})}\BibitemShut {NoStop}%
\bibitem [{\citenamefont {Gambino}\ \emph {et~al.}(2014)\citenamefont
  {Gambino}, \citenamefont {Lombardo},\ and\ \citenamefont
  {Sammartino}}]{gambino2014turing}%
  \BibitemOpen
  \bibfield  {author} {\bibinfo {author} {\bibfnamefont {G.}~\bibnamefont
  {Gambino}}, \bibinfo {author} {\bibfnamefont {M.}~\bibnamefont {Lombardo}},\
  and\ \bibinfo {author} {\bibfnamefont {M.}~\bibnamefont {Sammartino}},\
  }\href@noop {} {\bibfield  {journal} {\bibinfo  {journal} {Acta Appl. Math.}\
  }\textbf {\bibinfo {volume} {132}},\ \bibinfo {pages} {283} (\bibinfo {year}
  {2014})}\BibitemShut {NoStop}%
\bibitem [{\citenamefont {Boas}(2006)}]{boas2006mathematical}%
  \BibitemOpen
  \bibfield  {author} {\bibinfo {author} {\bibfnamefont {M.~L.}\ \bibnamefont
  {Boas}},\ }\href@noop {} {\emph {\bibinfo {title} {Mathematical Methods in
  the Physical Sciences}}},\ \bibinfo {edition} {3rd}\ ed.\ (\bibinfo
  {publisher} {Wiley},\ \bibinfo {address} {Hoboken, NJ},\ \bibinfo {year}
  {2006})\BibitemShut {NoStop}%
\bibitem [{\citenamefont {Ford}\ \emph {et~al.}(1991)\citenamefont {Ford},
  \citenamefont {Phillips}, \citenamefont {Quinn},\ and\ \citenamefont
  {Lauffenburger}}]{ford1991measurement}%
  \BibitemOpen
  \bibfield  {author} {\bibinfo {author} {\bibfnamefont {R.~M.}\ \bibnamefont
  {Ford}}, \bibinfo {author} {\bibfnamefont {B.~R.}\ \bibnamefont {Phillips}},
  \bibinfo {author} {\bibfnamefont {J.~A.}\ \bibnamefont {Quinn}},\ and\
  \bibinfo {author} {\bibfnamefont {D.~A.}\ \bibnamefont {Lauffenburger}},\
  }\href@noop {} {\bibfield  {journal} {\bibinfo  {journal} {Biotechnol.
  Bioeng.}\ }\textbf {\bibinfo {volume} {37}},\ \bibinfo {pages} {647}
  (\bibinfo {year} {1991})}\BibitemShut {NoStop}%
\bibitem [{\citenamefont {Saragosti}\ \emph {et~al.}(2011)\citenamefont
  {Saragosti}, \citenamefont {Calvez}, \citenamefont {Bournaveas},
  \citenamefont {Perthame}, \citenamefont {Buguin},\ and\ \citenamefont
  {Silberzan}}]{saragosti2011directional}%
  \BibitemOpen
  \bibfield  {author} {\bibinfo {author} {\bibfnamefont {J.}~\bibnamefont
  {Saragosti}}, \bibinfo {author} {\bibfnamefont {V.}~\bibnamefont {Calvez}},
  \bibinfo {author} {\bibfnamefont {N.}~\bibnamefont {Bournaveas}}, \bibinfo
  {author} {\bibfnamefont {B.}~\bibnamefont {Perthame}}, \bibinfo {author}
  {\bibfnamefont {A.}~\bibnamefont {Buguin}},\ and\ \bibinfo {author}
  {\bibfnamefont {P.}~\bibnamefont {Silberzan}},\ }\href@noop {} {\bibfield
  {journal} {\bibinfo  {journal} {Proc. Natl. Acad. Sci. U.S.A.}\ }\textbf
  {\bibinfo {volume} {108}},\ \bibinfo {pages} {16235} (\bibinfo {year}
  {2011})}\BibitemShut {NoStop}%
\bibitem [{\citenamefont {Newman}(2013)}]{newman2013computational}%
  \BibitemOpen
  \bibfield  {author} {\bibinfo {author} {\bibfnamefont {M.~E.~J.}\
  \bibnamefont {Newman}},\ }\href@noop {} {\emph {\bibinfo {title}
  {Computational Physics}}}\ (\bibinfo  {publisher} {CreateSpace Independent},\
  \bibinfo {address} {San Bernardino, CA},\ \bibinfo {year} {2013})\BibitemShut
  {NoStop}%
\bibitem [{\citenamefont {Sittner}\ \emph {et~al.}(2017)\citenamefont
  {Sittner}, \citenamefont {Bar-David}, \citenamefont {Glinert}, \citenamefont
  {Ben-Shmuel}, \citenamefont {Weiss}, \citenamefont {Schlomovitz},
  \citenamefont {Kobiler},\ and\ \citenamefont {Levy}}]{sittner2017pathology}%
  \BibitemOpen
  \bibfield  {author} {\bibinfo {author} {\bibfnamefont {A.}~\bibnamefont
  {Sittner}}, \bibinfo {author} {\bibfnamefont {E.}~\bibnamefont {Bar-David}},
  \bibinfo {author} {\bibfnamefont {I.}~\bibnamefont {Glinert}}, \bibinfo
  {author} {\bibfnamefont {A.}~\bibnamefont {Ben-Shmuel}}, \bibinfo {author}
  {\bibfnamefont {S.}~\bibnamefont {Weiss}}, \bibinfo {author} {\bibfnamefont
  {J.}~\bibnamefont {Schlomovitz}}, \bibinfo {author} {\bibfnamefont
  {D.}~\bibnamefont {Kobiler}},\ and\ \bibinfo {author} {\bibfnamefont
  {H.}~\bibnamefont {Levy}},\ }\href@noop {} {\bibfield  {journal} {\bibinfo
  {journal} {PloS one}\ }\textbf {\bibinfo {volume} {12}},\ \bibinfo {pages}
  {e0186613} (\bibinfo {year} {2017})}\BibitemShut {NoStop}%
\bibitem [{\citenamefont {Bonazzi}\ \emph {et~al.}(2018)\citenamefont
  {Bonazzi}, \citenamefont {Schiavo}, \citenamefont {Machata}, \citenamefont
  {Djafer-Cherif}, \citenamefont {Nivoit}, \citenamefont {Manriquez},
  \citenamefont {Tanimoto}, \citenamefont {Husson}, \citenamefont {Henry},
  \citenamefont {Chat{\'e}} \emph {et~al.}}]{bonazzi2018intermittent}%
  \BibitemOpen
  \bibfield  {author} {\bibinfo {author} {\bibfnamefont {D.}~\bibnamefont
  {Bonazzi}}, \bibinfo {author} {\bibfnamefont {V.~L.}\ \bibnamefont
  {Schiavo}}, \bibinfo {author} {\bibfnamefont {S.}~\bibnamefont {Machata}},
  \bibinfo {author} {\bibfnamefont {I.}~\bibnamefont {Djafer-Cherif}}, \bibinfo
  {author} {\bibfnamefont {P.}~\bibnamefont {Nivoit}}, \bibinfo {author}
  {\bibfnamefont {V.}~\bibnamefont {Manriquez}}, \bibinfo {author}
  {\bibfnamefont {H.}~\bibnamefont {Tanimoto}}, \bibinfo {author}
  {\bibfnamefont {J.}~\bibnamefont {Husson}}, \bibinfo {author} {\bibfnamefont
  {N.}~\bibnamefont {Henry}}, \bibinfo {author} {\bibfnamefont
  {H.}~\bibnamefont {Chat{\'e}}}, \emph {et~al.},\ }\href@noop {} {\bibfield
  {journal} {\bibinfo  {journal} {Cell}\ }\textbf {\bibinfo {volume} {174}},\
  \bibinfo {pages} {143} (\bibinfo {year} {2018})}\BibitemShut {NoStop}%
\bibitem [{\citenamefont {Bender}\ \emph {et~al.}(2013)\citenamefont {Bender},
  \citenamefont {Berntson}, \citenamefont {Parker},\ and\ \citenamefont
  {Samuel}}]{bender2013observation}%
  \BibitemOpen
  \bibfield  {author} {\bibinfo {author} {\bibfnamefont {C.~M.}\ \bibnamefont
  {Bender}}, \bibinfo {author} {\bibfnamefont {B.~K.}\ \bibnamefont
  {Berntson}}, \bibinfo {author} {\bibfnamefont {D.}~\bibnamefont {Parker}},\
  and\ \bibinfo {author} {\bibfnamefont {E.}~\bibnamefont {Samuel}},\
  }\href@noop {} {\bibfield  {journal} {\bibinfo  {journal} {Am. J. Phys.}\
  }\textbf {\bibinfo {volume} {81}},\ \bibinfo {pages} {173} (\bibinfo {year}
  {2013})}\BibitemShut {NoStop}%
\bibitem [{\citenamefont {Mathijssen}\ \emph
  {et~al.}(2018{\natexlab{a}})\citenamefont {Mathijssen}, \citenamefont
  {Guzm{\'a}n-Lastra}, \citenamefont {Kaiser},\ and\ \citenamefont
  {L{\"o}wen}}]{mathijssen2018nutrient}%
  \BibitemOpen
  \bibfield  {author} {\bibinfo {author} {\bibfnamefont {A.~J. T.~M.}\
  \bibnamefont {Mathijssen}}, \bibinfo {author} {\bibfnamefont
  {F.}~\bibnamefont {Guzm{\'a}n-Lastra}}, \bibinfo {author} {\bibfnamefont
  {A.}~\bibnamefont {Kaiser}},\ and\ \bibinfo {author} {\bibfnamefont
  {H.}~\bibnamefont {L{\"o}wen}},\ }\href@noop {} {\bibfield  {journal}
  {\bibinfo  {journal} {Phys. Rev. Lett.}\ }\textbf {\bibinfo {volume} {121}},\
  \bibinfo {pages} {248101} (\bibinfo {year} {2018}{\natexlab{a}})}\BibitemShut
  {NoStop}%
\bibitem [{\citenamefont {Mathijssen}\ \emph
  {et~al.}(2018{\natexlab{b}})\citenamefont {Mathijssen}, \citenamefont
  {Jeanneret},\ and\ \citenamefont {Polin}}]{mathijssen2018universal}%
  \BibitemOpen
  \bibfield  {author} {\bibinfo {author} {\bibfnamefont {A.~J. T.~M.}\
  \bibnamefont {Mathijssen}}, \bibinfo {author} {\bibfnamefont
  {R.}~\bibnamefont {Jeanneret}},\ and\ \bibinfo {author} {\bibfnamefont
  {M.}~\bibnamefont {Polin}},\ }\href@noop {} {\bibfield  {journal} {\bibinfo
  {journal} {Phys. Rev. Fluids}\ }\textbf {\bibinfo {volume} {3}},\ \bibinfo
  {pages} {033103} (\bibinfo {year} {2018}{\natexlab{b}})}\BibitemShut
  {NoStop}%
\end{thebibliography}
%
\end{document}